\documentclass[final,5p,times,twocolumn,nopreprintline]{elsarticle}
\usepackage[T1]{fontenc}

\usepackage{amsmath}
\usepackage{amsfonts}
\usepackage{amssymb}
\usepackage{physics}
\usepackage{slashed}
\usepackage{simplewick}
\usepackage{mathtools}
\usepackage{braket}
\usepackage{bm} 
\usepackage{dsfont}

\usepackage{siunitx}
\usepackage{orcidlink}
\usepackage{placeins}
\usepackage{graphicx}
\usepackage{float}
\usepackage[caption=false]{subfig}

\usepackage{listings}

\usepackage{hyperref}
\hypersetup{ 
    pdfnewwindow=true,      
    colorlinks=true,       
    allcolors=[RGB]{31 119 180}
} 
\usepackage{tikz}
\usetikzlibrary{calc}
\usetikzlibrary{patterns}
\usetikzlibrary{patterns}
\usetikzlibrary{positioning,decorations.pathmorphing,decorations.markings,snakes}

\usepackage{xcolor}
\definecolor{jpac-blue}{rgb}{0.12,0.47,0.71}
\definecolor{jpac-orange}{rgb}{1,0.5,0.05}
\definecolor{jpac-green}{rgb}{0,0.62,0.45}
\definecolor{jpac-red}{rgb}{0.84,0.15,0.15}
\definecolor{jpac-purple}{rgb}{0.58,0.40,0.71}
\definecolor{jpac-brown}{rgb}{0.54,0.34,0.29}
\definecolor{jpac-pink}{rgb}{0.89,0.47,0.76}
\definecolor{jpac-grey}{rgb}{0.5,0.5,0.5}
\definecolor{jpac-gold}{rgb}{0.74,0.74,0.13}
\definecolor{jpac-aqua}{rgb}{0.09,0.75,0.81}
\definecolor{jpac-white}{rgb}{1,1,1}

\usepackage[capitalise]{cleveref}

\usepackage{pgfplots,pgfplotstable}
\usetikzlibrary{pgfplots.groupplots}
\usetikzlibrary{decorations.markings}
\usepgfplotslibrary{fillbetween}
\pgfplotsset{compat=1.14}
\usepackage{tikz}
\usepackage{tikz-3dplot}
\usetikzlibrary{positioning}
\usetikzlibrary{calc}
\tikzset{->-/.style={decoration={
  markings,
  mark=at position .5 with {\arrow{>}}},postaction={decorate}}}
\tikzset{>=stealth}
\makeatletter
\tikzoption{canvas is plane}[]{\@setOxy#1}
\def\@setOxy O(#1,#2,#3)x(#4,#5,#6)y(#7,#8,#9)%
  {\def\tikz@plane@origin{\pgfpointxyz{#1}{#2}{#3}}%
   \def\tikz@plane@x{\pgfpointxyz{#4}{#5}{#6}}%
   \def\tikz@plane@y{\pgfpointxyz{#7}{#8}{#9}}%
   \tikz@canvas@is@plane
  }
\makeatother 

\newcommand\bsub{\begin{subequations}}
\newcommand\esub{\end{subequations}}

\usepackage{multirow}
\usepackage{dcolumn}
\usepackage{tabularx}

\usepackage{array}   
\newcolumntype{L}{>{$}l<{$}} 
\newcolumntype{R}{>{$}r<{$}}
\newcolumntype{C}{>{$}c<{$}}
\usepackage{xspace}
\usepackage[normalem]{ulem}
\usepackage[format=plain,justification=centering,singlelinecheck=false]{caption}
\usepackage[format=plain,justification=centering,singlelinecheck=false]{subcaption}
\usepackage{nameref}

\newcommand{\diff}{\ensuremath{\mathrm{d}}}
\newcommand{\mevnospace}{\ensuremath{{\mathrm{\,Me\kern -0.1em V}}}}
\newcommand{\gevnospace}{\ensuremath{{\mathrm{\,Ge\kern -0.1em V}}}}
\newcommand{\tevnospace}{\ensuremath{{\mathrm{\,Te\kern -0.1em V}}}}

\newcommand{\gev}{\gevnospace\xspace}

\newcommand{\gevsq}{\ensuremath{\gevnospace^2}\xspace}

\newcommand{\eg}{{\it e.g.}\xspace}
\newcommand{\cf}{{\it cf.}\xspace}

\newcommand{\ie}{{\it i.e.}\xspace}

\newcommand{\comment}[1]{}

\usepackage{todonotes}
\usepackage{enumitem}

\usepackage{fancyhdr}
\fancypagestyle{firstpage}{%
	
	\lhead{}
	\rhead{\small \begin{tabular}[t]{r}
        JLAB-THY-25-4567 \\
		MIT-CTP/5935
	\end{tabular}}
}

\journal{Physics Letters B}

\crefformat{appendix}{#2#1#3}

\begin{document}
\begin{frontmatter}

\title{
High-energy $\eta^{(\prime)}\pi$ photoproduction and the nature of exotic waves
}

\author[jlab,ub]{Gloria~Monta\~na\,\orcidlink{0000-0001-8093-6682} \corref{cor1}}
\ead{gmontana@jlab.org}
\cortext[cor1]{Corresponding author}
\author[ub]{Vincent~Mathieu\,\orcidlink{0000-0003-4955-3311}}
\author[ceem,indiana]{Vanamali~Shastry\,\orcidlink{0000-0003-1296-8468}}
\author[AGH]{{\L}ukasz~Bibrzycki\,\orcidlink{0000-0002-6117-4894}}
\author[uned]{C\'esar~Fern\'andez-Ram\'irez\,\orcidlink{0000-0001-8979-5660}}
\author[ub]{Nadine~Hammoud\,\orcidlink{0000-0002-8395-0647}}
\author[mit]{Robert~J.~Perry\,\orcidlink{0000-0002-2954-5050}}
\author[messina,catania]{Alessandro~Pilloni\,\orcidlink{0000-0003-4257-0928}}
\author[jlab,odu]{Arkaitz~Rodas\,\orcidlink{0000-0003-2702-5286}}
\author[wm,messina,ucb,lbnl]{Wyatt~A.~Smith\,\orcidlink{0009-0001-3244-6889}}
\author[jlab,ceem,indiana]{Adam~P.~Szczepaniak\,\orcidlink{0000-0002-4156-5492}}
\author[HISKP]{Daniel~Winney\,\orcidlink{0000-0002-8076-243X}}

\author{\\[.4cm] (Joint Physics Analysis Center)}

\affiliation[jlab]{organization={Theory Center, Thomas  Jefferson  National  Accelerator  Facility}, city={Newport  News}, state={VA}, postcode={23606}, country={USA}}
\affiliation[ub]{organization={Departament de F\'isica Qu\`antica i Astrof\'isica and Institut de Ci\`encies del Cosmos, Universitat de Barcelona}, postcode={E-08028}, city={Barcelona}, country={Spain}}
\affiliation[ceem]{organization={Center for Exploration of Energy and Matter, Indiana University}, city={Bloomington}, state={IN}, postcode={47403}, country={USA}}
\affiliation[indiana]{organization={Department of Physics, Indiana University}, city={Bloomington}, state={IN}, postcode={47405}, country={USA}}
\affiliation[AGH]{organization={AGH University of Krakow, Faculty of Physics and Applied Computer Science}, postcode={PL-30-059}, city={Krak\'ow}, country={Poland}}
\affiliation[uned]{organization={Departamento de F\'isica Interdisciplinar, Universidad Nacional de Educaci\'on a Distancia (UNED)}, postcode={E-28040}, city={Madrid}, country={Spain}}
\affiliation[mit]{organization={Center for Theoretical Physics -- a Leinweber Institute, Massachusetts Institute of Technology}, city={Cambridge}, state={MA}, postcode={02139}, country={USA}}
\affiliation[messina]{organization={Dipartimento di Scienze Matematiche e Informatiche, Scienze Fisiche e Scienze della Terra, Universit\`a degli Studi di Messina}, postcode={I-98166}, city={Messina}, country={Italy}}
\affiliation[catania]{organization={INFN Sezione di Catania}, postcode={I-95123}, city={Catania}, country={Italy}}
\affiliation[odu]{organization={Department of Physics, Old Dominion University}, city={Norfolk}, state={VA}, postcode={23529}, country={USA}}
\affiliation[wm]{organization={Department of Physics, William \& Mary},city={Williamsburg},state={VA},postcode={23187},country={USA}}
\affiliation[ucb]{organization={Department of Physics, University of California}, city={Berkeley}, state={CA}, postcode={94720}, country={USA}}
\affiliation[lbnl]{organization={Nuclear Science Division, Lawrence Berkeley National Laboratory}, city={Berkeley}, state={CA}, postcode={94720}, country={USA}}
\affiliation[HISKP]{organization={Helmholtz-Institut f\"{u}r Strahlen- und Kernphysik (Theorie) and Bethe Center for Theoretical Physics, Universit\"{a}t Bonn}, postcode={D-53115}, city={Bonn}, country={Germany}}



\begin{abstract}
The observation of hybrid mesons in photoproduction experiments can provide essential insight into the inner workings of quantum chromodynamics in the strong coupling regime. In particular, the study of final $\eta^{(\prime)}\pi$ states is of great interest due to the presence of the lowest lying hybrid candidate with manifestly exotic quantum numbers, the $\pi_1(1600)$. In this work, a double-vector exchange model with Reggeized $\rho$ and $\omega$ trajectories is developed to describe the photoproduction of $\eta^{(\prime)}\pi$  in the high-mass region. Results are presented for the differential cross sections and forward-backward asymmetries in the energy region of interest to the GlueX experiment. The model contains no free parameters, and reproduces the magnitude and $t$-dependence of existing CLAS data at \mbox{$E_\gamma=5$\gev.}
The model predicts a stronger asymmetry in the $\eta^\prime \pi$ channel than in the $\eta \pi$ channel, consistent with what has previously been observed in pion beam experiments.  This suggests that the sizeable production of exotic odd waves in $\eta'\pi$ is not necessarily related to the presence of gluon-rich environments.
Confirmation of these predicted asymmetries from forthcoming GlueX data would enable further predictions of the low-energy spectrum.
\end{abstract}

\end{frontmatter}

\thispagestyle{firstpage}

\section{Introduction}
\label{sec:introduction}

Attempts to understand exotic hadrons that are forbidden by the na\"ive quark model have been a main driving force in hadron spectroscopy in recent years, as such states are allowed in quantum chromodynamics (QCD). In particular, the search for hybrid mesons containing valence gluons was proposed as a tool to understand the features of gluodynamics in the nonperturbative regime~\cite{Meyer:2015eta,Gross:2022hyw}.
The most promising channels to look for these states have quantum numbers $J^{PC}=1^{-+}$, which are explicitly forbidden for a $q\bar q $ pair alone.

Evidence for a spin-exotic signal has been found by observing structures in odd partial waves in the $\eta^{(\prime)}\pi$ final states. Early experimental observations reported evidence for two possible states, the $\pi_1(1400)$ and the $\pi_1(1600)$, decaying into $\eta\pi$ and $\eta'\pi$, respectively~\cite{Aoyagi:1993kn, VES:1993scg, E852:1997gvf,E852:2001ikk}. Subsequent high-statistics measurements by the COMPASS experiment at CERN, based on diffractive production of $\eta^{(\prime)}\pi$ from a pion beam~\cite{COMPASS:2014vkj}, provided further evidence of exotic contributions in these channels. The presence of two nearby $1^{-+}$ states was puzzling.
Recent coupled-channels analyses provided evidence that both observations are consistent with a single resonance pole~\cite{JPAC:2018zyd,CrystalBarrel:2019zqh,Kopf:2020yoa}. A lattice QCD calculation at the $\textrm{SU}(3)_\textrm{F}$ symmetric point validated this result soon after~\cite{Woss:2020ayi}. Experimental evidence for an isoscalar partner in the $\eta\eta'$ final state has been found by \mbox{BESIII}~\cite{BESIII:2022riz}.

At $\eta^{(\prime)}\pi$ invariant masses above the resonance region, the process is dominated by Regge exchanges in the crossed channels. Analyticity and crossing symmetry imply that the low- and high-energy regions are connected, and knowledge about one region implies constraints on the other~\cite{Mathieu:2015gxa,JPAC:2016lnm,Mathieu:2017but}. In particular, a forward-backward angular asymmetry at high energies implies the presence of strong exotic partial waves at low energies. 

Indeed, data from the COMPASS experiment above the resonance region exhibits an asymmetry which further supports the identification of an exotic meson in the resonant region. This data can be understood in terms of Regge theory, where the interplay between different trajectories (the mesons $a_2$, $f_2$, and Pomeron) gives rise to the distinct patterns in the intensity distributions. In particular, a larger forward-backward asymmetry in the $\eta^\prime\pi$ final state as compared to $\eta\pi$ is understood within the Regge picture as being due to the larger Pomeron-$\eta^\prime$ coupling~\cite{Bibrzycki:2021rwh}. This hierarchy is natural if both the Pomeron and the $\eta^\prime$ contain a large gluonic component. Then, the connection between high-energy and low-energy scattering imposed by analyticity and crossing symmetry makes it plausible for the $\pi_1 (1600)$ to be prominent in the $\eta^\prime \pi$ lineshape due to its gluonic content.

The search for the $\pi_1(1600)$ and for other members of the lowest lying hybrid multiplet is the main goal of the GlueX experiment at Jefferson Lab, which utilizes a photon beam. Because of the photon's quantum numbers, Pomeron exchange is not allowed in $\eta^{(\prime)} \pi$ photoproduction, and it remains to be seen if the same argument about the connection between beam asymmetry and the nature of the $\pi_1$ holds. 
Upper limits on the $\pi_1(1600)$ photoproduction cross section and projections to the $\eta^{(\prime)}\pi$ channels have recently been estimated~\cite{GlueX:2024erj}, and the measurement of the forward-backward asymmetry at high energy is ongoing~\cite{RebecaPhD}. 
At GlueX energies, subleading Regge effects and contamination from nucleon resonances must be taken into account, which no theoretical model has fully considered so far.
The development of high-energy models is an essential first step towards implementing analyticity constraints in partial-wave extractions, which would allow a more robust determination of the $\pi_1$ resonance parameters. 

In this Letter we develop a high-energy model for $\eta^{(\prime)}\pi$ photoproduction, valid in the kinematics dominated by double-Regge exchanges. The model compares well with $\eta\pi$ CLAS data~\cite{CLAS:2020rdz}.
We compute the forward-backward asymmetry, finding a similar pattern to the pion beam data from \mbox{COMPASS}~\cite{COMPASS:2014vkj,Bibrzycki:2021rwh}. This provides further evidence that the $\eta^\prime \pi$ final state couples strongly to the exotic $1^{-+}$, even in the absence of Pomeron exchange. This suggests that the sizeable production of exotic odd waves in $\eta'\pi$ is not necessarily related to the presence of gluon-rich environments.
The Letter is organized as follows. In~\cref{sec:model} we introduce the fixed-spin double-vector exchange mechanism for $\eta^{(\prime)}\pi$ photoproduction. The Reggeization of this model to describe the double-Regge region is presented in~\cref{sec:reggeization}. Results for angular distributions, cross sections, and asymmetries are discussed in~\cref{sec:results}, followed by the conclusions in~\cref{sec:conclusions}.

\section{Double-vector exchange model}
\label{sec:model}
We consider the reaction 
\begin{align}\label{eq:reaction}
    \gamma(q_\gamma) + p(q_p) \to 
    \eta^{(\prime)}(q_\eta) + \pi (q_\pi) + p(q_{p'}) \ ,
\end{align}
in the Gottfried-Jackson (GJ) frame of the $\eta^{(\prime)}\pi$ system, in which the $z$-axis is aligned with the photon beam, the nucleons lie in the $xz$ plane, and both nucleon momenta have negative $x$ components (see \cref{sec:kinematics} for details). 

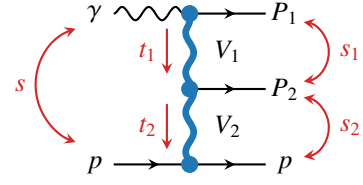
\begin{figure}
\centering
    \begin{tikzpicture}
        \draw[thick, -stealth] (0,0) -- (0.6,0);
        \draw[thick, -stealth] (0.5,0) -- (1.6,0);
        \draw[thick] (1.5,0) -- (2,0);
        \draw[decorate, thick, decoration=snake] (0,2) -- (1,2);
        \draw[thick, -stealth] (1,2) -- (1.6,2);
        \draw[thick] (1.5,2) -- (2,2);
         \draw[decorate,color=jpac-blue,line width=2.5pt,decoration={snake,amplitude=1.5,segment length=15}] (1,0) -- (1,2);
         \draw[thick, -stealth] (1,1) -- (1.6,1);
         \draw[thick] (1.5,1) -- (2,1);
         \filldraw[jpac-blue] (1,0) circle (3pt);
         \filldraw[jpac-blue] (1.,1) circle (3pt);
         \filldraw[jpac-blue] (1,2) circle (3pt);
         \node at (-0.25,2) {$\gamma$};
         \node at (-0.25,0) {$p$};
         \node at (2.25,2) {$P_1$};
         \node at (2.25,1) {$P_2$};
         \node at (2.25,0) {$p$};
         \node at (1.5,0.5) {$V_2$};
         \node at (1.5,1.5) {$V_1$};
         \draw[thick, -stealth, color=jpac-red] (0.7,1.8) -- (0.7,1.2) node[midway, left] {$t_1$};
         \draw[thick, -stealth, color=jpac-red] (0.7,0.8) -- (0.7,0.2) node[midway, left] {$t_2$};
         \draw[stealth-stealth, thick, color=jpac-red] (-0.5,1.8) arc (110:250:0.8) node[midway, left] {$s$};
         \draw[stealth-stealth, thick, color=jpac-red] (2.5,1.1) arc (-80:80:0.4) node[midway, right] {$s_1$};
         \draw[stealth-stealth, thick, color=jpac-red] (2.5,0.1) arc (-80:80:0.4) node[midway, right] {$s_2$};
    \end{tikzpicture}
\caption{Generic diagram for the photoproduction of two pseudoscalars $P_1$ and $P_2$ off a proton via the exchange of two vectors $V_1$ and $V_2$; $s$, $s_1$, $s_2$ denote squared center-of-mass energies of the subsystems, and $t_1$, $t_2$ the squared momentum transfers. }
\label{fig:diagram}
\end{figure}

As Reggeons emerge from the sum of definite spin-parity exchanges, we start by considering the values compatible with symmetries.
The simplest diagram of this form is the exchange of two vector mesons as shown in~\cref{fig:diagram}. Natural-parity exchanges are expected to provide the leading contribution 
at high energies. 
Because the meson produced at the photon vertex carries most of the beam momentum in the Lab frame, we will denote ``fast-$\eta$'' (``fast-$\pi$'') the diagrams in which $P_1$ is the $\eta^{(\prime)}$ ($\pi$) meson, and $P_2$ the other pseudoscalar meson. $V_1$ and $V_2$ are the top and the bottom exchanged vectors, respectively. Four configurations are considered: 
\bsub\begin{align}\label{eq:diagrams-fast-pi}
    \text{fast-}\pi &: (V_1,V_2) = \left\{(\rho,\rho) ;\, (\omega, \omega)\right\}\, ,
    \\ \label{eq:diagrams-fast-eta}
    \text{fast-}\eta &: (V_1,V_2) = \left\{(\rho,\omega) ;\, (\omega, \rho)\right\}\, .
\end{align}\esub
The $\phi$ exchange is also allowed, but it would be responsible for significant differences in the $\eta$ and $\eta'$ beam asymmetries, which are not supported by existing measurements~\cite{Mathieu:2017jjs, GlueX:2019adl}. 
We have explicitly checked that including the $\phi$ exchange in our calculation would modify the cross section by only a few percent, well below other model uncertainties. Hence, its contribution is neglected.

Each contribution to the amplitude factorizes naturally into three vertices, whose Lorentz structures are uniquely determined by the quantum numbers of the participating particles.
The top vertex involves the vector-vector-pseudoscalar interaction
\begin{align}\label{eq:gammaVP}
\left \langle P_1 V_1| \gamma  \right\rangle & = \frac{g_{\gamma V_1 P_1}}{m_0} i \epsilon_{\alpha \beta \mu\nu} \varepsilon^{\alpha}_{\gamma} q^\beta_{\gamma} q_{V_1}^\mu \varepsilon^{\nu*}_{V_1} \ ,
\end{align}
where $q_{i}$ and $\varepsilon_{i}^\nu$ denote momenta and polarization vectors of the photon and the exchanged vector meson. 
Similarly, the middle vertex is given by
\begin{align}\label{eq:VVP}
\left \langle P_2 V_2| V_1 \right\rangle & = \frac{g_{V_1 V_2 P_2}}{m_0} i \epsilon_{\alpha \beta \mu\nu} \varepsilon^{\alpha}_{V_1} q^\beta_{V_1} q_{V_2}^\mu \varepsilon^{\nu*}_{V_2} \ ,
\end{align}
and the bottom vertex describes the interaction of a vector meson with the nucleons
\begin{align}\label{eq:VNN}
\langle p|V_2 p \rangle=\varepsilon^{\nu}_{V_2}\,\bar u(q_{p'})\left[ (g^{V_2}_1 + g^{V_2}_2) \gamma_{ \nu} - g^{V_2}_2 \frac{(q_p+q_{p'})_{\nu}}{2m_p} \right] u(q_p) \ .
\end{align}

The magnitudes of the couplings are listed in~\cref{tab:couplings-all}, where we have set the reference scale $m_0=1\gev$. The photon couplings $g_{\gamma V_1P_1}$ are extracted from the measured radiative decay of the $\omega$ and $\rho$ mesons into $\gamma\pi^0$ and $\gamma\eta$, while for the heavier $\eta'$ pseudoscalar meson, we use its radiative decay into $\gamma\omega$ and $\gamma\rho^0$~\cite{ParticleDataGroup:2024cfk}.
The couplings to the nucleons $g_1^{V_2}$ and $g_2^{V_2}$ are taken from the phenomenological Regge residues~\cite{Irving:1977ea}.
The middle-vertex couplings $g_{V_1V_2P_2}$ are determined from the \mbox{$\omega\to3\pi$} decay, assuming it is saturated by an intermediate $\rho$, \ie \mbox{$\omega\to\rho\pi\to 3\pi$}~\cite{Gell-Mann:1962hpq}. The remaining couplings are obtained through $\textrm{SU}(3)_\textrm{F}$ symmetry and $\eta-\eta'$ mixing. Further details are provided in~\cref{sec:couplings}.

\begin{table}
\caption{Vertex couplings (with $m_0=1\gev$).}
\label{tab:couplings-all}
\setlength{\tabcolsep}{10pt}
\centering
\begin{tabular}{cc|cc|cc}
        \hline 
        \multicolumn{2}{c|}{$g_{\gamma V_1P_1}/m_0$ } & \multicolumn{2}{c|}{$g_{V_1V_2P_2}/m_0$ } & \multicolumn{2}{c}{$g_{1,2}^{V_2}$ } \\
        \multicolumn{2}{c|}{ [\cf~\cref{eq:gammaVP}]} & \multicolumn{2}{c|}{ [\cf~\cref{eq:VVP}]} & \multicolumn{2}{c}{[\cf~\cref{eq:VNN}]} \\
        \hline
        $g_{\gamma\omega\pi}$ & 0.701 & $g_{\omega\rho\pi}$ & 14.27 & $g^{\omega}_1$ & 5.75 \\
        $g_{\gamma\omega\eta}$ & 0.135 & $\phi_\eta$ & $41.4^\circ$ & $g^{\omega}_2$ & 1.31 \\
        $g_{\gamma\omega\eta^\prime}$ & 0.127 & $g_{\omega\omega\eta}$ & 10.70 & $g^{\rho}_1$ & 1.15 \\
        $g_{\gamma\rho\pi}$ & 0.223 & $g_{\omega\omega\eta^\prime}$ & 9.44 & $g^{\rho}_2$ & 9.2 \\
        $g_{\gamma\rho\eta}$ & 0.480 & $g_{\rho\rho\eta}$ &  10.70 & & \\
        $g_{\gamma\rho\eta^\prime}$ & 0.398 & $g_{\rho\rho\eta^\prime}$ & 9.44 & & \\
        \hline
    \end{tabular}
\end{table}

The unpolarized differential cross section depends on five independent kinematic variables, for which we choose the total center-of-mass energy squared, $s=(q_\gamma+q_p)^2$; the invariant mass squared of the $\eta^{(\prime)}\pi$ pair, $s_1\equiv s_{\eta\pi}=(q_\eta+q_\pi)^2$; the momentum transfer from the target to the recoil proton, $t_{pp} = (q_p - q_{p'})^2$; and the two angles $\Omega_{\text{GJ}}\equiv(\theta,\phi)$, defined as the polar and azimuthal angles of the $\eta^{(\prime)}$ momentum in the GJ frame. With this definition, forward ($\cos \theta \simeq 1$) events correspond to fast-$\eta$, and backward ($\cos \theta \simeq -1$) events to fast-$\pi$.
In terms of these variables, the differential cross section is given by
\begin{align}\label{eq:diffcrosssec}
\frac{\diff^4 \sigma}{\diff^2 \Omega_{\text{GJ}}\diff t_{pp}\diff s_{\eta\pi}}  &= \frac{1}{16(2\pi)^4} \frac{\lambda^\frac12(s,s_{\eta\pi}, m_p^2)}{ s_{\eta\pi}\,\lambda(s,0, m_p^2)} \,  \frac{1}{4}
\sum_{\lambda_\gamma,\lambda,\lambda'} | A_{\lambda_\gamma \lambda \lambda'}|^2 \ ,
\end{align}
where the amplitude $A_{\lambda_\gamma\lambda\lambda'}$ is constructed from the product of the three vertices defined in~\cref{eq:gammaVP,eq:VVP,eq:VNN} and the two intermediate vector propagators, and depends on the helicities of the photon, target and recoil, denoted by $\lambda_\gamma$, $\lambda$, and $\lambda'$, respectively. It takes the form
\begin{align} \label{eq:model}
A_{\lambda_\gamma \lambda \lambda'} & = 
 \sum_{\text{diagrams}} \left( \frac{g_{\gamma V_1P_1}}{m_0}  \frac{g_{V_1 V_2 P_2}}{m_0} \right) 
 \frac{ N_{\lambda_\gamma \lambda \lambda'} }{(m_{V_1}^2 - t_{1})(m_{V_2}^2 - t_2)} \ ,
\end{align}
where $t_{1}=(q_\gamma-q_{P_1})^2$ and $t_2\equiv t_{pp}$ are the squared momentum transfers at the top and bottom vertices, and the spinor structure is encapsulated in the numerator
\begin{align}\label{eq:KasAB}
N_{\lambda_\gamma \lambda \lambda'} 
& \equiv 
  \bar u_{\lambda'} \left[- g^{V_2}_2 A_{\lambda_\gamma}+ (g^{V_2}_1 + g^{V_2}_2) B_{\lambda_\gamma}^\mu \gamma_\mu\right] u_\lambda \ .
\end{align}
Using $q_{P_1}$ and $q_{P_2}$ to denote the momenta of the two pseudoscalar mesons, the Lorentz tensors $A_{\lambda_\gamma}$ and $B_{\lambda_\gamma}^{\bar\nu}$ are defined as
\bsub\begin{align}\label{eq:A}
    A_{\lambda_\gamma} & = \varepsilon_{\alpha \beta \mu\nu} \epsilon^\alpha_{\lambda_\gamma} q^\beta_\gamma q_{P_1}^\mu \varepsilon^{\nu \bar \beta \bar \mu\bar \nu} q_{P_2,\,\bar \beta}(q_\gamma-q_{P_1})_{\bar \mu} \frac{(q_p+q_{p'})_{\bar \nu}}{2m_p} \ ,
\\ \label{eq:B}
    B^{\bar \nu}_{\lambda_\gamma} & = \varepsilon_{\alpha \beta \mu\nu} \epsilon^\alpha_{\lambda_\gamma} q^\beta_\gamma q_{P_1}^\mu \varepsilon^{\nu \bar \beta \bar \mu\bar \nu} q_{P_2,\,\bar \beta}(q_\gamma-q_{P_1})_{\bar \mu} \ .
\end{align}\esub
The numerator can be split into 
\begin{align}
    N_{\lambda_\gamma \lambda \lambda'} & = G_{\lambda_\gamma \Lambda} K_{\lambda_\gamma \Lambda}\ . 
\end{align}
The helicity transfer between the nucleons is defined as \mbox{$\Lambda \equiv \lambda'-\lambda$}. 
The $G_{\lambda_\gamma \Lambda}$ factor contains the couplings and spin-flip factors, and is given by 
\begin{align}
    G_{\lambda_\gamma \Lambda} & = \,g_2^{V_2} \frac{|\bm p_p'|^2}{m_p^2}\,\delta_{\Lambda,0} + (g_1^{V_2} + g_2^{V_2})\left( \delta_{\Lambda,0} -  \sqrt{2} \frac{E_p'}{m_p}\, \delta_{|\Lambda|,1}\right) \ ,
\end{align}
where ${E_p'=\sqrt{t_{pp}}/2}$ and $|\bm p_p'|=\lambda^\frac12(t_{pp},m^2_p,m^2_p)/2\sqrt{t_{pp}}$ are the energy and magnitude of the three-momentum of the target proton in the $V_2$ rest frame. 
Importantly, the model contains no free parameters: all couplings entering the calculation are fixed from experimental data (see \cref{sec:couplings}).

The $K_{\lambda_\gamma \Lambda}$ factor is purely kinematical, and it is most conveniently evaluated in the $t$-channel frame by exploiting its factorized structure:~\cref{eq:A,eq:B} are computed in the rest frame of the exchanged meson $V_1$, while the nucleon spinors are evaluated in the $V_2$ rest frame. The orientation of these frames and the respective kinematics are defined in \cref{sec:kinematics}. This yields a particularly simple expression for $K_{\lambda_\gamma\Lambda}$, 
\begin{align}\label{eq:Kfactor-tchannel}
    K_{\lambda_\gamma\Lambda} = 2m_p t_1|\bm p_{P_1}||\bm p_{P_2}| \lambda_\gamma \sum_{\xi=-1}^1(-\xi)e^{-i\xi\omega}d^1_{\lambda_\gamma\xi}(\theta_1)\,d^1_{\xi\Lambda}(\theta_2) \ ,
\end{align}
where ${\bm p_{P_1}}$ and $\bm p_{P_2}$ are the three-momenta of the pseudoscalar mesons in the $V_1$ rest frame. 

When evaluating the equations for a given diagram, one needs to perform the following substitution.
For the fast-$\eta$ diagrams, $q_{P_1}\equiv q_\eta$, $q_{P_2}\equiv q_\pi$, while for the fast-$\pi$ diagrams, $q_{P_1}\equiv q_\pi$, $q_{P_2}\equiv q_\eta$.  
The GJ angles are related to $t_1$ and $s_2$, corresponding to $t_{\gamma\eta}=(q_\gamma-q_\eta)^2$ and $s_{\pi p}=(q_\pi+q_{p'})^2$ for fast-$\eta$ amplitudes, and $t_{\gamma\pi}=(q_\gamma-q_\pi)^2$ and $s_{\eta p}=(q_\eta+q_{p'})^2$ for fast-$\pi$ amplitudes. In \cref{sec:kinematics}, we specify the relevant kinematic quantities for the fast-$\eta$ diagrams. 

\begin{figure*}[htb]
\begin{center}
\includegraphics[width=0.95\linewidth]{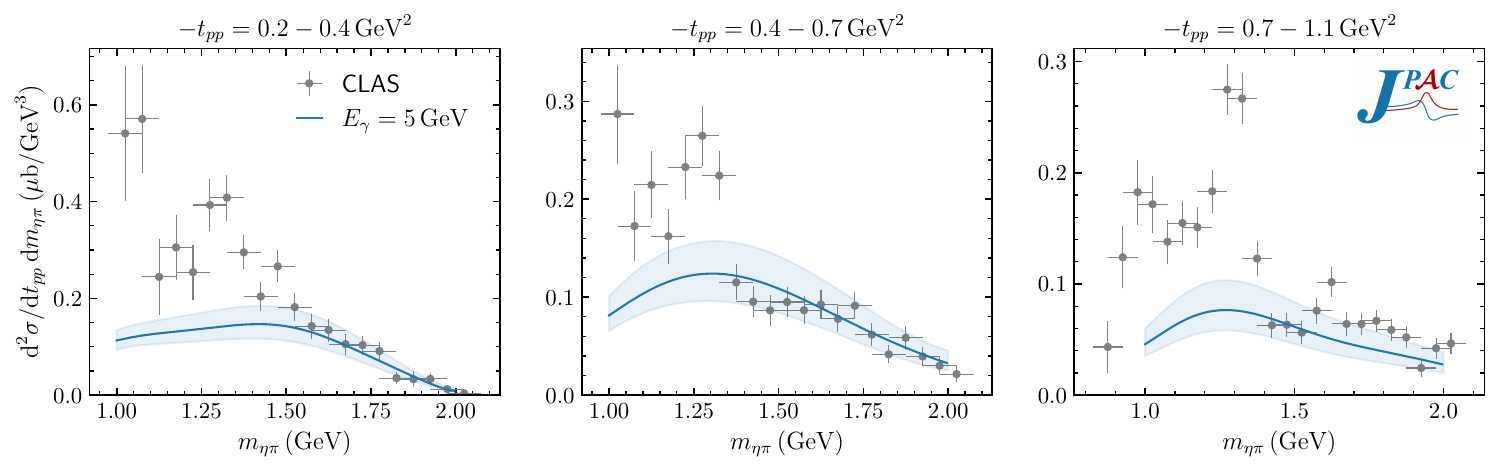}
\end{center}
\caption{Differential cross section as a function of the $m_{\eta\pi} = \sqrt{s_{\eta\pi}}$ invariant mass for $E_\gamma=5\gev$, in slices of $t_{pp}$. Shaded bands represent the model uncertainty obtained by varying $s_0$ in the range $0.95-1.05\gevsq$. Experimental data are from~\cite{CLAS:2020rdz}.}
\label{fig:dsigmaClasV}
\end{figure*}

\section{Reggeization}
\label{sec:reggeization}
With the double-vector exchange model of~\cref{sec:model}, we establish the vertex structures and the couplings. The latter are determined from data. A realistic description of the high-energy regime of interest is obtained only after Reggeization, which requires summing over the full tower of higher spin resonances appearing in the $\rho$ and $\omega$ Regge trajectories.

To enforce the analytic behavior expected from the double-Regge exchange formalism~\cite{Brower:1974yv,Shimada:1978sx}, the amplitude of~\cref{eq:model} is multiplied by a Regge factor of the form 
\begin{align}
    &R(\alpha_1,\alpha_2,s_{1},s_{2})=(\alpha_1-1)\Gamma(1-\alpha_1)(\alpha_2-1)\Gamma(1-\alpha_2) \nonumber \\
    &\quad \times\left[ \xi_1\xi_{21}\kappa^{1-\alpha_1}V(\alpha_1,\alpha_2,\kappa)+\xi_2\xi_{12}\kappa^{1-\alpha_2}V(\alpha_2,\alpha_1,\kappa)\right] \nonumber \\
    &\quad \times\left(\frac{s_{1}}{s_0}\right)^{\alpha_1-1}\left(\frac{s_{2}}{s_0}\right)^{\alpha_2-1} \ ,
    \label{eq:ReggeFactor}
\end{align}
where $\alpha_i$ are Regge trajectories, $\kappa^{-1}\equiv s/(\alpha' s_1s_2)$, and $V(\alpha_1, \alpha_2, \kappa)$ may be interpreted as a vertex function, whose explicit form is given below. Thus, after Reggeization, the numerator of each diagram is replaced by
\begin{align}\label{eq:ReggeNumerator}
    N_{\lambda_\gamma\lambda\lambda'} \to R (\alpha_1,\alpha_2,s_{1},s_{2}) \times N_{\lambda_\gamma\lambda\lambda'}\ .
\end{align}
The factors $(\alpha_i-1)$ in~\cref{eq:ReggeFactor} cancel the poles of the vector-meson propagators in~\cref{eq:model} as $\alpha_i\to1$, thereby effectively replacing them by Regge propagators. The scale parameter $s_0$ is conventionally chosen as $1\gevsq$.

The Regge trajectories are taken to be linear
\begin{align}\label{eq:ReggeTrajectory}
    \alpha_i(t_i)=1 + \alpha' (t_i - m_{V_i}^2)\ ,
\end{align}
with nearly degenerate $\rho$ and $\omega$ trajectories sharing a common slope $\alpha'=0.9\gev^{-2}$. 

The numerator $RN_{\lambda_\gamma\lambda\lambda'}$ in~\cref{eq:ReggeNumerator} has poles at positive odd integer values of the trajectories $\alpha_1$ and $\alpha_2$, corresponding to the spins of the physical particles exchanged in the $t$ channel. The residue at the vector poles ($\alpha_1,\alpha_2=1$) is normalized to unity. The signature factors 
\begin{align}
    \xi_i&=\frac{1}{2}\left(1-e^{-i\pi\alpha_i}\right) \ , &
    \xi_{ij}&=\frac{1}{2}\left(1+e^{-i\pi(\alpha_i-\alpha_j)}\right)\ ,
\end{align}
remove the even signature poles, ensuring that only the odd signature $(-1)^J=-1$ poles couple to $\gamma P_1$ in the top vertex, and that the middle vertex is likewise restricted to odd-signature trajectories.

The vertex function $V(\alpha_1,\alpha_2,\kappa)$ is analytic in all of its variables. It represents a sum over infinitely many Reggeon-Reggeon-particle couplings in the middle vertex and reduces to a polynomial in $\kappa$ for integer values of $\alpha_1$ and $\alpha_2$~\cite{Drummond:1969jv}. We use the particularly compact form obtained by assuming, for simplicity, that all these couplings are identical~\cite{Shimada:1978sx},
\begin{align}
    \label{eq:RR_vertex}
    V(\alpha_1,\alpha_2,\kappa)=\frac{\Gamma(\alpha_1-\alpha_2)}{\Gamma(1-\alpha_2)}\,\phantom{}_1F_1(1-\alpha_1,1-\alpha_1+\alpha_2,-\kappa) \ ,
\end{align}
where $\phantom{}_1F_1$ is the confluent hypergeometric function of the first kind. The form of the middle vertex in~\cref{eq:RR_vertex} can also be shown to emerge in the double-Regge limit of the scalar five-particle dual-resonance amplitude~\cite{Brower:1974yv,Shi:2014nea}.

The double-Regge limit is defined by taking $s, s_1,s_2\to \infty$ for fixed $t_1$, $t_2$, while keeping $\kappa$ and the ratio $s_1/s_2$ finite.  
In this limit, the kinematical factor associated with the double-vector exchange model is
\begin{align}
    K_{\lambda_\gamma\Lambda} &\to \sqrt{2}m_p\frac{t_1}{4|\bm p_{P_2}'||\bm p_p'|}s_{1}s_{2}\sin\omega \ ,
\end{align}
which implies $K_{\lambda_\gamma\Lambda} \propto s$ since $\sin \omega$ is finite in this limit. Here, ${\bm p_p'}$ and $\bm p_{P_2}'$ are the three-momenta of the recoil proton
and the $P_2$ meson, respectively, in the $V_2$ rest frame, and are functions of momentum transfers.  The calculation is detailed in \cref{sec:kinematicsingularities}.
The two terms in Eq.~\eqref{eq:ReggeFactor} scale as $s^{\alpha_{1,2}-1}s_{2,1}^{\alpha_{2,1}-\alpha_{1,2}}$, so that the Reggeized amplitude exhibits the correct asymptotic behavior, $A\sim s^{\alpha_{1,2}}s_{2,1}^{\alpha_{2,1}-\alpha_{1,2}}$
in the double-Regge region, with no simultaneous singularities in overlapping channels. 

\section{Results}
\label{sec:results}
Before presenting numerical predictions for cross sections and asymmetries, it is instructive to comment on the overall impact of Reggeization. The non-Reggeized double-vector exchange model of~\cref{sec:model}, where $\rho$ and $\omega$ are treated as fixed-spin exchanges, leads to cross sections that are unrealistically large and inconsistent with existing photoproduction data. 
This is expected, since the pointlike exchange picture neglects the high-energy analytic constraints enforced by Regge theory. In particular, the vector-exchange amplitudes grow approximately linearly with $s_{1,2}$, violating subchannel unitarity.

This is fixed by embedding the $\rho$ and $\omega$ exchanges in their full Regge trajectories according to~\cref{sec:reggeization}, which dramatically reduces the magnitude of the amplitude.
The large number of crossed-channel partial waves interferes constructively for small transferred momenta and destructively at large angles, resulting in the well-known forward and backward peaks characteristic of high-energy reactions. 
Across the relevant kinematic range, the Reggeized amplitudes are suppressed by more than one order of magnitude, bringing the resulting cross section into a reasonable scale for $\eta^{(\prime)}\pi$ photoproduction. The suppression is primarily driven by the $t_{1,2}$-dependence of the Regge trajectories $\alpha_{1,2}(t_{1,2})$ in the Regge factor in~\cref{eq:ReggeFactor}, which become increasingly negative for $t_{pp}<0$. We note that, while the dimensional scale parameter $s_0\sim1\gevsq$ cannot be predicted and effectively serves as a free parameter (see, \eg, Ref.~\cite{JointPhysicsAnalysisCenter:2024kck}), here we adopt its standard phenomenological value rather than fitting it. To estimate the model dependence associated with this choice, we vary $s_0$ within a narrow interval around its nominal value. This variation effectively captures uncertainties related to slopes of the Regge trajectory and the contribution from subdominant exchanges, while the vertex couplings of the double-vector model are well constrained and their uncertainties have negligible impact on the observables considered here.

In the following, we focus on the Reggeized model. To compare with experimental measurements from CLAS~\cite{CLAS:2020rdz}, the differential cross section in~\cref{eq:diffcrosssec} must be integrated over the GJ angles:
\begin{align}
\frac{\diff^2 \sigma}{\diff t_{pp} \diff m_{\eta\pi}} &= 2m_{\eta\pi}\int \diff^2 \Omega_\text{GJ} 
\frac{\diff^4 \sigma}{\diff^2 \Omega_\text{GJ} \diff t_{pp}
\diff s_{\eta\pi}} \ .
\end{align}

A direct comparison with data is shown in~\cref{fig:dsigmaClasV}. The  Reggeization procedure brings the results to overlap with the CLAS measurement at large invariant masses. The central choice $s_0=1\gevsq$ provides the best description, and the variation from $0.95$ to $1.05\gevsq$ remains compatible with the data, defining a reasonable range to estimate the model uncertainty. At lower $m_{\eta\pi}$, however, where $s_{\eta \pi}$ resonances appear, the Regge approach breaks and is not expected to provide a good description. 

\begin{figure}[htb]
\begin{center}
\includegraphics[width=0.95\linewidth]{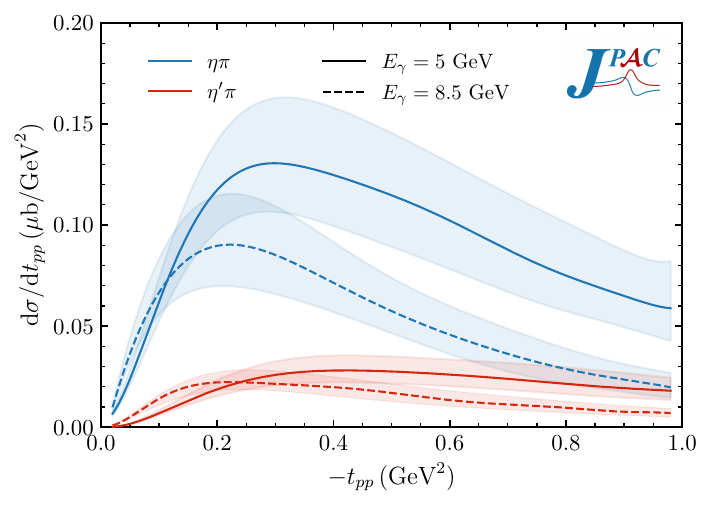}
\end{center}
\caption{Differential cross section for the photoproduction of $\eta\pi$ and $\eta'\pi$  as a function of $-t_{pp}$ for values of the beam energy corresponding to the CLAS and GlueX experiments. Uncertainties are computed as in~\cref{fig:dsigmaClasV}.}
\label{fig:dsigmavst}
\end{figure}

The dependence on the momentum transfer $-t_{pp}$ is illustrated in~\cref{fig:dsigmavst}. At both energies, the cross section falls steeply with increasing $-t_{pp}$, as expected for processes dominated by Regge exchanges. The comparison shows the larger suppression at the higher GlueX energy, consistent with the $s$-dependence of the Reggeized amplitude. The variation of $s_0$ mainly produces a nearly uniform shift in the overall normalization of the cross section. We note that, while the present uncertainty estimates are modest, this work aims to establish a consistent framework for $\eta^{(\prime)}\pi$ photoproduction in the double-Regge region which can be systematically refined once high-statistics GlueX data becomes available.

\subsection{Angular distribution}
\begin{figure}[htb]
\centering
\includegraphics[width=\linewidth]{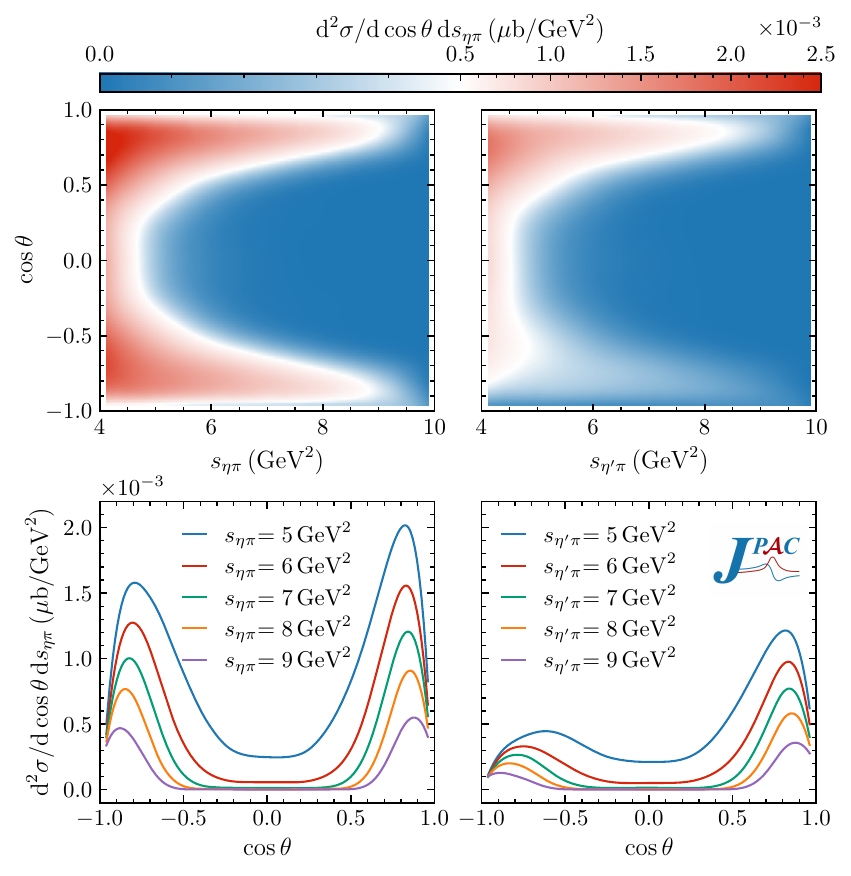}
\caption{Angular distribution of the differential cross section as a function of the $\eta\pi$ (left) and $\eta'\pi$ (right) invariant mass squared. Top: two-dimensional distribution; bottom: corresponding slices at fixed $s_{\eta^{(\prime)}\pi}$.}
\label{fig:dsigma2d}
\end{figure}

Now we present the angular distribution in the GJ frame. In the top panel of~\cref{fig:dsigma2d}, the differential cross section is shown as a 2D distribution in the polar angle $\theta$ and invariant mass squared, for both the $\eta\pi$ (left) and the $\eta'\pi$ (right) final states. Both distributions exhibit two prominent peaks in the forward and backward directions, which gradually fade with increasing $s_{\eta\pi}$, a behavior characteristic of the double-Regge region.

The bottom panels display slices of these distributions, illustrating the evolution of the structures with $s_{\eta\pi}$. For the $\eta'\pi$ channel, the asymmetry between the forward and backward peaks is more pronounced than in the $\eta\pi$ case, as we discuss in the next Section. 

\subsection{Forward-backward asymmetry}
The forward-backward asymmetry provides a measure of the angular dependence of the reaction and is defined as
\begin{equation} \label{eq:asym}
A_{\textrm{FB}}(s_{\eta\pi})   = 
\frac{F(s_{\eta\pi}) - B(s_{\eta\pi})}{F(s_{\eta\pi}) + B(s_{\eta\pi})} \ ,
\end{equation} 
where $F(s_{\eta\pi})$ and $B(s_{\eta\pi})$ are the forward and backward differential cross sections
\bsub
\begin{align}  
F(s_{\eta\pi}) &= \int_0^1 \diff\!\cos\theta \int \diff\phi\, \diff t_{pp} \frac{\diff ^4\sigma}{\diff^2\Omega_{\text{GJ}} \diff t_{pp} \diff s_{\eta\pi}} \ ,\\
B(s_{\eta\pi}) &= \int_{-1}^0 \diff\!\cos\theta \int \diff\phi\, \diff t_{pp} \frac{\diff ^4\sigma}{\diff^2\Omega_{\text{GJ}} \diff t_{pp} \diff s_{\eta\pi}} \ .
\end{align} \esub
We do not display uncertainty bands for $A_{\textrm{FB}}$, as the dependence on $s_0$ cancels almost entirely between the forward and backward contributions, making the resulting variation too small to be visible.

Previous studies of this quantity at COMPASS find that the asymmetries are both negative, with that of $\eta'\pi$ being larger in magnitude than that of $\eta\pi$. The sign can be understood as being due to the contribution from the gluon-rich Pomeron in the top exchange, which gives a larger contribution to fast-$\pi$ diagrams.

\begin{figure}[htb]
\begin{center}
\includegraphics[width=\linewidth]{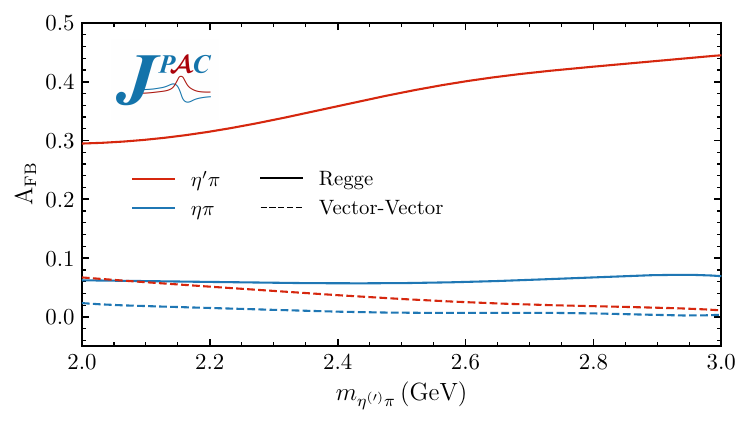}
\end{center}
\caption{Forward-backward asymmetry as a function of $\eta^{(\prime)}\pi$ invariant mass.}
\label{fig:asym}
\end{figure}

As shown in~\cref{fig:asym}, the forward-backward asymmetry predicted by the double-vector exchange model without Reggeization is relatively small for both $\eta\pi$ and $\eta'\pi$, consistent with the nearly flat angular distribution of the corresponding differential cross section. 
The inclusion of the Regge factor~\cref{eq:ReggeFactor} generates the characteristic forward and backward peaks, and leads to a significantly larger forward-backward asymmetry, particularly for the $\eta'\pi$ channel. In the absence of Pomeron exchange, the stronger enhancement observed for $\eta'\pi$ does not follow trivially and could originate from two effects: 
\begin{enumerate}
\item[(i)]
The different coupling strengths of the $\rho$ and $\omega$ trajectories to the $\eta$ and $\eta'$, which modify the relative weights of the contributing amplitudes. This turns out to be irrelevant, as the couplings are of the same order of magnitude, and no large variation to the asymmetry occurs if one swaps the sets of couplings.
\item[(ii)]
The $t_{1,2}$-dependence of the Regge trajectories, which alters the interference patterns as the reaction kinematics shifts with the heavier $\eta'$. This makes the larger contribution to the asymmetry, as one can see by continuously varying the mass of the $\eta'$ to match the mass of the $\eta$.
\end{enumerate}
These possible mechanisms suggest that the sizeable production of exotic odd waves in $\eta'\pi$ is not necessarily related to the presence of a gluon-rich environment such as the one related to Pomeron exchange.

In experimental analyses, the forward and backward differential cross sections are often obtained after applying kinematical cuts to suppress backgrounds from baryon-resonance production, such as the $\Delta(1232)$. This is particularly important at GlueX, where such backgrounds can be sizeable and cuts on kinematical quantities are required to isolate the double-Regge region. To enable a meaningful comparison with upcoming GlueX measurements, in~\cref{sec:asym-cuts} we compute the forward-backward asymmetry with kinematical cuts. 

\section{Conclusions}
\label{sec:conclusions}
The photoproduction of $\eta\pi$ and $\eta'\pi$ in the double-Regge region has been investigated using a model based on Reggeized double-vector exchanges. All couplings to the $\rho$ and $\omega$ are determined from independent measurements, and the analytic structure dictated by Regge theory is fully incorporated. The resulting differential cross sections agree with the magnitude and $t$-dependence observed in CLAS data at $E_\gamma=5\gev$ and provide concrete predictions at higher photon energies relevant for upcoming GlueX analyses.  

The angular distributions exhibit the forward and backward peaks, characteristic of the double-Regge regime. In comparison with pion-induced reactions, the absence of Pomeron exchange leads to opposite-sign forward-backward asymmetries. Interestingly, the $\eta'\pi$ asymmetry is larger than the $\eta\pi$, similarly to what was observed by COMPASS with pion beams. This suggests that the sizeable production of exotic odd waves in $\eta'\pi$ is not necessarily related to the presence of a gluon-rich environment such as the one related to Pomeron exchange, and therefore that the strong coupling of $\eta'\pi$ to the hybrid $\pi_1(1600)$ may manifest clearly in photoproduction, providing further support for the $\pi_1(1600)$ searches in this channel at GlueX.

The present framework also lays the groundwork for a future study of finite-energy sum rules. By integrating the high-energy amplitudes, it is possible to connect the double-Regge asymptotics to the contributions of low-energy resonances in the direct channel. This will allow a quantitative test of duality and provide additional constraints on partial-wave extraction in the resonance-dominated regions.

\section*{Acknowledgements}
This work was supported by the U.S. Department of Energy contract \mbox{DE-AC05-06OR23177}, under which Jefferson Science Associates, LLC operates Jefferson Lab,
by U.S. Department of Energy Grant Nos.~\mbox{DE-FG02-87ER40365}, 
and \mbox{DE-SC0011090},
and it contributes to the aims of the U.S. Department of Energy \mbox{ExoHad} Topical Collaboration, contract \mbox{DE-SC0023598}.	
RJP acknowledges support by the Simons Foundation award Simons Collaboration on Confinement and \mbox{QCD StringsMPS-QCD-00994314} and by MIT.
GM, VM and NH have been supported by the projects \mbox{CEX2024-001451-M} (Unidad de Excelencia ``María de Maeztu''), \mbox{PID2020-118758GB-I00},
and GM additionally by \mbox{PID2023-147112NB-C21}, 
all financed by \mbox{MICIU/AEI/10.13039/501100011033/} and FEDER, UE, as well as by the EU \mbox{STRONG-2020} project, under the program \mbox{H2020-INFRAIA-2018-1} Grant Agreement \mbox{No.~824093}.
VM is a Professor Serra H\'unter. VM and NH acknowledge support from \mbox{CNS2022-136085}. GM was additionally supported by the Beatriu de Pinós program by AGAUR, Grant \mbox{No. BP 2024 00189}.

\appendix
\section{Kinematics}\label{sec:kinematics}
The process
\begin{align}\label{eq:reaction_app}
    \gamma(q_\gamma) + p(q_p) \to 
    \eta^{(\prime)}(q_\eta) + \pi^0 (q_\pi) + p(q_{p'}) \ ,
\end{align}
requires five independent variables. Our cross section~\cref{eq:diffcrosssec} was expressed in terms of $s$, $s_{\eta\pi}$, $t_{pp}$ and $\Omega_{\text{GJ}} = (\theta,\phi)$, defined in the main text. In this Section, we provide their expression in terms of Lorentz-invariant variables. 
For the fast-$\eta$ diagrams, we conveniently choose the following set of independent Mandelstam variables: 
\begin{align}\nonumber
    s & = (q_\gamma + q_p)^2, & 
    s_{\eta\pi} & = (q_\eta + q_\pi)^2, & 
    s_{\pi p} & = (q_\pi + q_p)^2,
    \\
    t_{\gamma \eta} & = (q_\gamma - q_\eta)^2, &
    t_{pp} & = (q_p - q_{p'})^2\ .
\end{align}
For fast-$\pi$ one needs the obvious replacements $\pi \leftrightarrow \eta$.
Any other Lorentz-invariant quantity can be expressed in terms of the five Mandelstam variables above. For instance, the following relations will be useful:
\bsub\begin{align}
    t_{\gamma p}&= (q_\gamma-q_{p'})^2 = s_{\eta \pi}-s-t_{pp}+2m_p^2 \ , \\
    t_{\gamma\pi}&= (q_\gamma-q_\pi)^2 = t_{pp}-t_{\gamma\eta}-s_{\eta \pi}+m_\eta^2+m_\pi^2 \ , 
    \\
    t_{p\pi} &= (q_p-q_\pi)^2 = t_{\gamma \eta} - t_{pp} - s_{\pi p} + m_\pi^2 + 2 m_p^2 \ ,
    \\
    s_{\eta p}&= (q_\eta+q_p)^2=s-s_{\eta\pi}-s_{\pi p}+m_\eta^2+m_\pi^2+m_p^2 \ .
\end{align}\esub

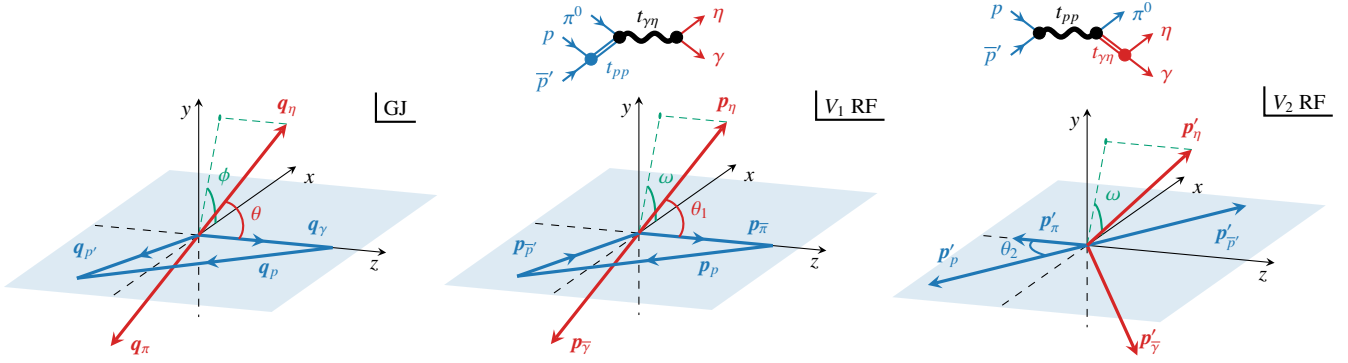
\begin{figure*}
\tdplotsetmaincoords{75}{20}
\begin{tikzpicture}[scale=0.75, every node/.style={scale=0.8},tdplot_main_coords]
\coordinate (origin) at (0,0,0);
\pgfmathsetmacro{\lx}{5}
\pgfmathsetmacro{\ly}{2.5}
\pgfmathsetmacro{\lz}{3.5}
\pgfmathsetmacro{\nlx}{-4.5}
\pgfmathsetmacro{\nly}{-1.5}
\pgfmathsetmacro{\nlz}{-2}
\pgfmathsetmacro{\ax}{4}
\pgfmathsetmacro{\az}{3}
\pgfmathsetmacro{\nax}{-4.5}
\pgfmathsetmacro{\naz}{-2}
\pgfmathsetmacro\pe{ 2.5}
\pgfmathsetmacro\posi{0.05}
\fill[jpac-blue, opacity=0.15] (\az,\ax,0) -- (\naz,\ax,0) -- (\naz,\nax,0) -- (\az,\nax,0) -- cycle; 
    \draw[->] (0,0,0) -- (\lz,0,0) node[anchor=north east]{$z$};
    \draw[->] (0,0,0) -- (0,\lx,0) node[anchor=north west]{$x$};
    \draw[->] (0,0,0) -- (0,0,\ly) node[anchor=north east]{$y$};
    \draw[dashed] (0,0,0) -- (\nlz,0,0) node[anchor=north east]{};
    \draw[dashed] (0,0,0) -- (0,\nlx,0) node[anchor=north west]{};
    \draw[dashed] (0,0,0) -- (0,0,\nly) node[anchor=south]{};
\pgfmathsetmacro\th{60}
\pgfmathsetmacro\ph{30}
\pgfmathsetmacro\fpe{0.4}
\pgfmathsetmacro{\peZu}{cos(\th)};
\pgfmathsetmacro{\peXu}{sin(\th)*sin(\ph)};
\pgfmathsetmacro{\peYu}{sin(\th)*cos(\ph)};
\pgfmathsetmacro{\peXYu}{sin(abs(\th))};
\pgfmathsetmacro{\peZ}{\pe*\peZu};
\pgfmathsetmacro{\peX}{\pe*\peXu};
\pgfmathsetmacro{\peY}{\pe*\peYu};
\pgfmathsetmacro{\peXY}{\pe*sqrt(\peXu*\peXu + \peYu*\peYu)};
\pgfmathsetmacro{\peYZ}{\pe*sqrt(\peYu*\peYu + \peZu*\peZu)};
\pgfmathsetmacro{\peXZ}{\pe*sqrt(\peXu*\peXu + \peZu*\peZu)};
\coordinate (eta)  at ({\peZ},{+\peX},{+\peY});
\coordinate (pion)   at ({-\peZ},{-\peX},{-\peY});
\coordinate (etaY)   at (\peZ,\peX,0);
\coordinate (etaX)   at (\peZ,0,\peY);
\coordinate (etaZ)   at (0,\peX,\peY);
\pgfmathsetmacro\pg{ 4}
\pgfmathsetmacro\prix{4}
\pgfmathsetmacro\priy{-2}
\coordinate (gamma) at (+2.5,0,0);
\coordinate (target)   at (-1,-3.5,0);    
\begin{scope}[canvas is yz plane at x=0.0]
\draw[jpac-green,fill] (etaZ)+(\posi,0) arc (0:360:\posi) {};
\draw[jpac-green, thick] ({\fpe*\peXY},0.0) arc (0:90-\ph:{\fpe*\peXY}) node[near end,above right] {$\phi$};
\end{scope}
\draw[thin,color=jpac-green,densely dashed]  (eta) -- (etaZ)   node[below] {};    
\draw[thin,color=jpac-green,densely dashed]  (origin) -- (etaZ)   node[below] {};    
\draw[very thick,jpac-red,->]  (origin) -- (pion)   node[very near end, below right] {$\bm q_{\pi}$};
\draw[very thick,color=jpac-blue,->-] (origin) -- (gamma) node[near end,above right] {$\bm q_{\gamma}$}; 
\draw[very thick,color=jpac-blue,->-] (gamma) -- (target)    node[near start,below]  {$\bm q_p$};
\draw[very thick,color=jpac-blue,->-] (origin) -- (target)   node[near end,above left] {$\bm q_{ p'}$};
\draw[very thick,jpac-red,->]  (origin) -- (eta)   node[anchor=south] {$\bm q_\eta$}; 
\draw[jpac-red, thick] (0.3*\pe,0,0) to[bend right=\th] node[near start, above right] {$\theta$} ({0.3*\peZ},{0.3*\peX},{0.3*\peY});
\tdplotsetrotatedcoords{20}{0}{0}
\coordinate (Shift) at (-4,4,0.7);
\tdplotsetrotatedcoordsorigin{(Shift)}
\draw[tdplot_rotated_coords,thick] (5.5,2) node[below right] {GJ} -- (5.5,0) -- (6.2,0) ;
\end{tikzpicture}
\begin{tikzpicture}[scale=0.75, every node/.style={scale=0.8},tdplot_main_coords]
\coordinate (origin) at (0,0,0);
\pgfmathsetmacro{\lx}{5}
\pgfmathsetmacro{\ly}{2.5}
\pgfmathsetmacro{\lz}{3.5}
\pgfmathsetmacro{\nlx}{-4.5}
\pgfmathsetmacro{\nly}{-1.5}
\pgfmathsetmacro{\nlz}{-2}
\pgfmathsetmacro{\ax}{4}
\pgfmathsetmacro{\az}{3}
\pgfmathsetmacro{\nax}{-4.5}
\pgfmathsetmacro{\naz}{-2}
\pgfmathsetmacro\pe{ 2.5}
\pgfmathsetmacro\posi{0.05}
\fill[jpac-blue, opacity=0.15] (\az,\ax,0) -- (\naz,\ax,0) -- (\naz,\nax,0) -- (\az,\nax,0) -- cycle; 
    \draw[->] (0,0,0) -- (\lz,0,0) node[anchor=north east]{$z$};
    \draw[->] (0,0,0) -- (0,\lx,0) node[anchor=north west]{$x$};
    \draw[->] (0,0,0) -- (0,0,\ly) node[anchor=north east]{$y$};
    \draw[dashed] (0,0,0) -- (\nlz,0,0) node[anchor=north east]{};
    \draw[dashed] (0,0,0) -- (0,\nlx,0) node[anchor=north west]{};
    \draw[dashed] (0,0,0) -- (0,0,\nly) node[anchor=south]{};
\pgfmathsetmacro\th{60}
\pgfmathsetmacro\ph{30}
\pgfmathsetmacro\fpe{0.4}
\pgfmathsetmacro{\peZu}{cos(\th)};
\pgfmathsetmacro{\peXu}{sin(\th)*sin(\ph)};
\pgfmathsetmacro{\peYu}{sin(\th)*cos(\ph)};
\pgfmathsetmacro{\peXYu}{sin(abs(\th))};
\pgfmathsetmacro{\peZ}{\pe*\peZu};
\pgfmathsetmacro{\peX}{\pe*\peXu};
\pgfmathsetmacro{\peY}{\pe*\peYu};
\pgfmathsetmacro{\peXY}{\pe*sqrt(\peXu*\peXu + \peYu*\peYu)};
\pgfmathsetmacro{\peYZ}{\pe*sqrt(\peYu*\peYu + \peZu*\peZu)};
\pgfmathsetmacro{\peXZ}{\pe*sqrt(\peXu*\peXu + \peZu*\peZu)};
\coordinate (eta)  at ({\peZ},{+\peX},{+\peY});
\coordinate (gamma)   at ({-\peZ},{-\peX},{-\peY});
\coordinate (etaY)   at (\peZ,\peX,0);
\coordinate (etaX)   at (\peZ,0,\peY);
\coordinate (etaZ)   at (0,\peX,\peY);
\pgfmathsetmacro\pg{ 4}
\pgfmathsetmacro\prix{4}
\pgfmathsetmacro\priy{-2}
\coordinate (pion) at (+2.5,0,0);
\coordinate (target)   at (-1,-3.5,0);    
\begin{scope}[canvas is yz plane at x=0.0]
\draw[jpac-green,fill] (etaZ)+(\posi,0) arc (0:360:\posi) {};
\draw[jpac-green, thick] ({\fpe*\peXY},0.0) arc (0:90-\ph:{\fpe*\peXY}) node[near end,above right] {$\omega$};
\end{scope}
\draw[thin,color=jpac-green,densely dashed]  (eta) -- (etaZ)   node[below] {};    
\draw[thin,color=jpac-green,densely dashed]  (origin) -- (etaZ)   node[below] {};    
\draw[very thick,jpac-red,->]  (origin) -- (gamma)   node[very near end, below right] {$\bm p_{\overline\gamma}$};
\draw[very thick,color=jpac-blue,->-] (origin) -- (pion) node[near end,above right] {$\bm p_{\overline\pi}$}; 
\draw[very thick,color=jpac-blue,->-] (pion) -- (target)    node[near start,below]  {$\bm p_p$};
\draw[very thick,color=jpac-blue,->-] (target) -- (origin)   node[near start,above left] {$\bm p_{\overline p'}$};
\draw[very thick,jpac-red,->]  (origin) -- (eta)   node[anchor=south] {$\bm p_\eta$}; 
\draw[jpac-red, thick] (0.3*\pe,0,0) to[bend right=\th] node[near start, above right] {$\theta_1$} ({0.3*\peZ},{0.3*\peX},{0.3*\peY});
\tdplotsetrotatedcoords{20}{0}{0}
\coordinate (Shift) at (-4,10,0.7);
\tdplotsetrotatedcoordsorigin{(Shift)}
\draw[tdplot_rotated_coords,decorate, decorate,line width=2pt,decoration={snake,amplitude=1.5}] (0,0) -- (1,0) node[midway, above] {$t_{\gamma\eta}$};
\draw[tdplot_rotated_coords,thick, ->,jpac-red] (1,0) -- (1.5,1.5) node[right] {$\eta$};
\draw[tdplot_rotated_coords,thick, ->,jpac-red] (1,0) -- (1.5,-1.5) node[right] {$\gamma$};
\draw[tdplot_rotated_coords,thick, double, jpac-blue] (-0.5,-1.5) -- (0,0) node[near start, below right] {$t_{pp}$};
\draw[tdplot_rotated_coords,thick, ->-, jpac-blue]  (-0.5,1.5) -- (0,0) node[at start, left] {$\pi^0$};
\draw[tdplot_rotated_coords,thick, ->-, jpac-blue]  (-1,0) -- (-0.5,-1.5) node[at start, left] {$p$};
\draw[tdplot_rotated_coords,thick, ->-, jpac-blue]  (-1,-3) -- (-0.5,-1.5) node[at start,  left] {$\overline p'$};
\filldraw[tdplot_rotated_coords] (0,0) circle (3pt);
\filldraw[tdplot_rotated_coords] (1,0) circle (3pt);
\filldraw[tdplot_rotated_coords,jpac-blue] (-0.5,-1.5) circle (3pt);
\tdplotsetrotatedcoords{20}{0}{0}
\coordinate (Shift) at (-4,4,0.7);
\tdplotsetrotatedcoordsorigin{(Shift)}
\draw[tdplot_rotated_coords,thick] (5.5,2) node[below right] {$V_1$ RF} -- (5.5,0) -- (6.7,0) ;
\end{tikzpicture}
\begin{tikzpicture}[scale=0.75, every node/.style={scale=0.8},tdplot_main_coords]
\coordinate (origin) at (0,0,0);
\pgfmathsetmacro{\lx}{5}
\pgfmathsetmacro{\ly}{2.5}
\pgfmathsetmacro{\lz}{3.5}
\pgfmathsetmacro{\nlx}{-4.5}
\pgfmathsetmacro{\nly}{-1.5}
\pgfmathsetmacro{\nlz}{-2}
\pgfmathsetmacro{\ax}{4}
\pgfmathsetmacro{\az}{3}
\pgfmathsetmacro{\nax}{-4.5}
\pgfmathsetmacro{\naz}{-2}
\pgfmathsetmacro\pe{ 2.5}
\pgfmathsetmacro\posi{0.05}
\fill[jpac-blue, opacity=0.15] (\az,\ax,0) -- (\naz,\ax,0) -- (\naz,\nax,0) -- (\az,\nax,0) -- cycle; 
    \draw[->] (0,0,0) -- (\lz,0,0) node[anchor=north east]{$z$};
    \draw[->] (0,0,0) -- (0,\lx,0) node[anchor=north west]{$x$};
    \draw[->] (0,0,0) -- (0,0,\ly) node[anchor=north east]{$y$};
    \draw[dashed] (0,0,0) -- (\nlz,0,0) node[anchor=north east]{};
    \draw[dashed] (0,0,0) -- (0,\nlx,0) node[anchor=north west]{};
    \draw[dashed] (0,0,0) -- (0,0,\nly) node[anchor=south]{};
\pgfmathsetmacro\th{50}
\pgfmathsetmacro\ph{30}
\pgfmathsetmacro\fpe{0.4}
\pgfmathsetmacro{\peZu}{cos(\th)};
\pgfmathsetmacro{\peXu}{sin(\th)*sin(\ph)};
\pgfmathsetmacro{\peYu}{sin(\th)*cos(\ph)};
\pgfmathsetmacro{\peXYu}{sin(abs(\th))};
\pgfmathsetmacro{\peZ}{\pe*\peZu};
\pgfmathsetmacro{\peX}{\pe*\peXu};
\pgfmathsetmacro{\peY}{\pe*\peYu};
\pgfmathsetmacro{\peXY}{\pe*sqrt(\peXu*\peXu + \peYu*\peYu)};
\pgfmathsetmacro{\peYZ}{\pe*sqrt(\peYu*\peYu + \peZu*\peZu)};
\pgfmathsetmacro{\peXZ}{\pe*sqrt(\peXu*\peXu + \peZu*\peZu)};
\coordinate (eta)  at ({\peZ},{+\peX},{+\peY});
\coordinate (gamma)   at ({0.8*\peZ},{-\peX},{-\peY});
\coordinate (etaY)   at (\peZ,\peX,0);
\coordinate (etaX)   at (\peZ,0,\peY);
\coordinate (etaZ)   at (0,\peX,\peY);
\pgfmathsetmacro\pg{ 4}
\pgfmathsetmacro\prix{4}
\pgfmathsetmacro\priy{-2}
\coordinate (pion) at (-1.4,0,0);
\coordinate (target)   at (-1.7,-3.5,0);    
\coordinate (recoil)   at (1.7,3.5,0);    
\begin{scope}[canvas is yz plane at x=0.0]
\draw[jpac-green,fill] (etaZ)+(\posi,0) arc (0:360:\posi) {};
\draw[jpac-green, thick] ({\fpe*\peXY},0.0) arc (0:90-\ph:{\fpe*\peXY}) node[near end,above right] {$\omega$};
\end{scope}
\begin{scope}[canvas is xy plane at z=0.0]
\draw[jpac-blue, thick] (-1,0) arc (-180:-180+64:1) node[xshift=-6mm, yshift=1mm] {$\theta_2$};
\end{scope}
\draw[thin,color=jpac-green,densely dashed]  (eta) -- (etaZ)   node[below] {};    
\draw[thin,color=jpac-green,densely dashed]  (origin) -- (etaZ)   node[below] {};    
\draw[very thick,jpac-red,->]  (origin) -- (gamma)   node[very near end, right] {$\bm p'_{\overline\gamma}$};
\draw[very thick,color=jpac-blue,->] (origin) -- (pion) node[near end,above right] {$\bm p'_{\pi}$}; 
\draw[very thick,color=jpac-blue,->] (origin) -- (target)  node[near end,above left]  {$\bm p'_p$};
\draw[very thick,color=jpac-blue,->] (origin) -- (recoil)   node[near end,below right] {$\bm p'_{\overline p'}$};
\draw[very thick,jpac-red,->]  (origin) -- (eta)   node[anchor=south] {$\bm p'_\eta$}; 
\tdplotsetrotatedcoords{20}{0}{0}
\coordinate (Shift) at (-4,10,1);
\tdplotsetrotatedcoordsorigin{(Shift)}
\draw[tdplot_rotated_coords,decorate, decorate,line width=2pt,decoration={snake,amplitude=1.5}] (-0.5,0) -- (0.5,0) node[midway, above] {$t_{pp}$};
\draw[tdplot_rotated_coords,thick, ->,jpac-red] (1,-1.5) -- (1.5,0) node[right] {$\eta$};
\draw[tdplot_rotated_coords,thick, ->,jpac-red] (1,-1.5) -- (1.5,-3) node[right] {$\gamma$};
\draw[tdplot_rotated_coords,thick, double, jpac-red] (0.5,0) -- (1,-1.5) node[near start, below] {$t_{\gamma \eta}$};
\draw[tdplot_rotated_coords,thick, ->, jpac-blue]  (0.5,0) -- (1,1.5) node[right] {$\pi^0$};
\draw[tdplot_rotated_coords,thick, ->-, jpac-blue]  (-1,1.5) -- (-0.5,0) node[at start, left] {$p$};
\draw[tdplot_rotated_coords,thick, ->-, jpac-blue]  (-1,-1.5) -- (-0.5,0) node[at start,  left] {$\overline p'$};
\filldraw[tdplot_rotated_coords] (-0.5,0) circle (3pt);
\filldraw[tdplot_rotated_coords] (0.5,0) circle (3pt);
\filldraw[tdplot_rotated_coords,jpac-red] (1,-1.5) circle (3pt);
\tdplotsetrotatedcoords{20}{0}{0}
\coordinate (Shift) at (-4,4,1);
\tdplotsetrotatedcoordsorigin{(Shift)}
\draw[tdplot_rotated_coords,thick] (5.5,2) node[below right] {$V_2$ RF} -- (5.5,0) -- (6.7,0) ;
\end{tikzpicture}
\caption{Momenta in the $\eta\pi$ GJ frame (left), in the rest frame of the $V_1$ exchange (center) and of the $V_2$ exchange (right). The vectors in blue lie in the $xz$ plane.
}\label{fig:frames}
\end{figure*}

Experimentally, the kinematics of the reaction in~\cref{eq:reaction_app} are conveniently described in the GJ frame, as illustrated in the left plot in~\cref{fig:frames}. In this frame, the four-momenta are given by
\bsub\begin{align}
    q_\eta&=(\mathcal{E}_\eta,{\bm q_\eta}) \ , & q_\pi&=(\mathcal{E}_\pi,{\bm q_\pi}) \ , \\
    q_p&=(\mathcal{E}_p,{\bm q_p}) \ ,  & q_{p'}&=(\mathcal{E}_{p'},{\bm q_{p'}}) \ , \\
    q_\gamma&=(\mathcal{E}_\gamma,{\bm q_\gamma}) \ , 
\end{align}\esub
with ${\bm q_\pi}=-{\bm q_\eta}$. The beam direction $\bm q_\gamma$ defines the $z$-axis, and the $y$-axis is taken perpendicular to the production plane and oriented along ${\bm q_p} \times {\bm q_\gamma}$.

The energies can be expressed in terms of Mandelstam variables as 
\bsub\begin{align}
    \mathcal{E}_{\eta}&=(s_{\eta\pi}+m_\eta^2-m_\pi^2)/2\sqrt{s_{\eta\pi}} \ , \\
    \mathcal{E}_{\pi}&=(s_{\eta\pi}+m_\pi^2-m_\eta^2)/2\sqrt{s_{\eta\pi}}  \ , \\  
    \mathcal{E}_p&=(s+t_{pp}-m_p^2)/2\sqrt{s_{\eta\pi}} \ ,  \\
    \mathcal{E}_{p'}&=(s-s_{\eta\pi}-m_p^2)/2\sqrt{s_{\eta\pi}} \ ,  \\
    \mathcal{E}_\gamma&=(s_{\eta\pi}-t_{pp})/2\sqrt{s_{\eta\pi}} \ .
\end{align}\esub
The corresponding magnitude of the momenta are given by
\bsub\begin{align}
    |\bm q_\eta|&=|\bm q_\pi|=\lambda^{1/2}(s_{\eta\pi}, m_\eta^2, m_\pi^2)/2\sqrt{s_{\eta\pi}} \ , \\
    |\bm q_p|&=\lambda^{1/2}(s_{\eta\pi}, m_p^2, t_{\gamma p})/2\sqrt{s_{\eta\pi}} \ , \\
    |\bm q_{p'}|&=\lambda^{1/2}(s_{\eta\pi}, m_p^2, s)/2\sqrt{s_{\eta\pi}} \ , \\
    |\bm q_\gamma|&=\lambda^{1/2}(s_{\eta\pi}, 0, t_{pp})/2\sqrt{s_{\eta\pi}} \ .
\end{align}\esub

The angles between the beam and the target ($\xi$) and between the beam and the recoil proton ($\epsilon$) are given by
\bsub\begin{align}
    2 |\bm q_\gamma| |\bm q_p| \cos\xi &=  s- 2 \mathcal{E}_\gamma \mathcal{E}_p -m_p^2 \ , \\
    2 |\bm q_\gamma| |\bm q_{p'}| \cos\epsilon & = s- 2 \mathcal{E}_\gamma \mathcal{E}_{p'} -m_p^2- s_{\eta\pi} + t_{pp} \ . 
\end{align}\esub

Since the energies and angles $\xi$ and $\epsilon$ depend only on $s$, $s_{\eta\pi}$ and $t_{pp}$, the remaining angular variables $\theta$ and $\phi$ must depend on the other two independent invariants, $t_{\gamma\eta}$ and $s_{\pi p}$. Indeed, we find
\bsub\begin{align}
t_{\gamma\eta} &= m_\eta^2 - 2 \mathcal{E}_\gamma \mathcal{E}_\eta + 2 |\bm q_\gamma| |\bm q_\eta|\cos\theta \\
s_{\pi p} &= m_\pi^2+m_p^2 + 2 \mathcal{E}_\pi \mathcal{E}_{p'}  \\ \nonumber
&- 2 |\bm q_\eta| |\bm q_{p'}|\left(\sin \epsilon \sin\theta \cos\phi + \cos\epsilon \cos \theta \right) \ .
\end{align}\esub

The relevant frames for Reggeization are the rest frames of the $V_1$ and $V_2$ exchanges. The orientation of these frames is depicted in~\cref{fig:frames} (center and right plots), for the fast-$\eta$ case. We provide in the following paragraphs the relevant kinematic quantities corresponding to the fast-$\eta$ diagrams. For the fast-$\pi$ diagrams, energies, momenta, and angles need to be adapted accordingly.

The rest frame of $V_1$ corresponds to the rest frame of the crossed reaction
\begin{align}\label{eq:reactionV1}
    p(p_p, \nu) + \overline{p}(p_{\overline{p}'}, \nu') +{\pi}^0 (p_{\overline{\pi}}) \to {\gamma}(p_{\overline{\gamma}}, \nu_{\overline{\gamma}}) + 
    \eta^{(\prime)}(p_\eta) \ ,
\end{align}
where the crossed momenta are defined as \mbox{$p_{\overline{p}'}=-p_{p'}$}, \mbox{$p_{\overline{\pi}}=-p_{\pi}$}, and \mbox{$p_{\overline{\gamma}}=-p_{\gamma}$}. In this frame, the $z$-axis is taken along the direction of ${\bm p}_{\overline{\pi}}$, and the $y$-axis is defined along ${\bm p}_p \times {\bm p}_{\overline \pi}$. The four-momenta are given by
\bsub\begin{align}
    p_\eta&=(E_\eta,{\bm p_\eta}) \ , & p_{\overline{\pi}}&=(E_{\overline{\pi}},{\bm p_{\overline{\pi}}}) \ , \\
    p_p&=(E_p,{\bm p_p}) \ ,  & p_{\overline{p}'}&=(E_{\overline{p}'},{\bm p_{\overline{p}'}}) \ , \\
    p_{\overline{\gamma}}&=(E_{\overline{\gamma}},{\bm p_{\overline{\gamma}}}) \ ,
\end{align}\esub
with the corresponding energies
\bsub\begin{align}
    E_{\eta}&=(t_{\gamma\eta}+m_\eta^2)/2\sqrt{t_{\gamma\eta}} \ , \\
    E_{\overline{\pi}}&=(t_{\gamma\eta}+m_\pi^2-t_{pp})/2\sqrt{t_{\gamma\eta}} \ , \\ 
    E_p&=(t_{\gamma\eta}+m_p^2-s_{\pi p})/2\sqrt{t_{\gamma\eta}} \ , \\
    E_{\overline{p}'}&=(t_{\gamma \eta}+m_p^2-t_{p\pi})/2\sqrt{t_{\gamma\eta}} \ , \\
    E_{\overline{\gamma}}&=(t_{\gamma\eta}-m_\eta^2)/2\sqrt{t_{\gamma\eta}} \ .
\end{align}\esub
The momenta in the $V_1$ rest frame are given by
\bsub\begin{align}
    |\bm p_\eta| & = \lambda^{1/2}(t_{\gamma\eta}, 0, m_\eta^2) / 2\sqrt{t_{\gamma\eta}}\ ,\\
    |\bm p_{\bar \pi}| & = \lambda^{1/2}(t_{\gamma\eta}, t_{pp}, m_\pi^2) / 2\sqrt{t_{\gamma\eta}}\ ,\\
    |\bm p_{p}| & = \lambda^{1/2}(t_{\gamma\eta}, s_{\pi p}, m_p^2) / 2\sqrt{t_{\gamma\eta}}\ , \\
    |\bm p_{\bar p'}| & = \lambda^{1/2}(t_{\gamma\eta}, t_{p\pi}, m_p^2) / 2\sqrt{t_{\gamma\eta}}\ ,
\end{align}\esub
and $|\bm p_{\bar \gamma}| =  |\bm p_\eta|$ for the fast-$\eta$ diagrams. 

The scattering angle $\theta_1$ corresponds to the angle between the $\bm p_\eta$ and $\bm p_{\overline{\pi}}$ three-momenta.  We define as $\xi_1$ and $\epsilon_1$ the angles between $\bm p_p$ and $-\bm p_{\bar\pi}$, and $\bm p_{\bar p'}$ and $\bm p_{\bar \pi}$, respectively. These angles can be expressed in terms of invariants as follows
\begin{align}
\nonumber
\cos\theta_1 & = \frac{2 t_{\gamma\eta} (s_{\eta\pi}-m_\eta^2-m_\pi^2) + (t_{\gamma\eta}+m_\eta^2)(t_{\gamma\eta}+m_\pi^2-t_{pp})}
{ \lambda^{1/2}(t_{\gamma\eta}, m_\eta^2, 0) \lambda^{1/2}(t_{\gamma\eta},m_\pi^2,t_{pp})} \ ,
\\
\nonumber
\cos\epsilon_1 & =\frac{2 t_{\gamma\eta} (m_\pi^2+m_p^2-s_{\pi p}) + (t_{\gamma\eta}+m_\pi^2-t_{pp})(t_{\gamma\eta}+m_p^2-t_{p\pi})}
{ \lambda^{1/2}(t_{\gamma\eta}, m_\pi^2, t_{pp}) \lambda^{1/2}(t_{\gamma\eta},m_p^2,t_{p\pi})}\ ,
\\
\cos\chi_1 & = \frac{2 t_{\gamma\eta} (t_{p \pi} -m_p^2-m_\pi^2) - (t_{\gamma\eta}+m_\pi^2-t_{pp})(t_{\gamma\eta}+m_p^2-s_{\pi p})}
{ \lambda^{1/2}(t_{\gamma\eta},m_\pi^2,t_{pp}) \lambda^{1/2}(t_{\gamma\eta}, m_p^2, s_{\pi p})} \ .
\end{align}
These satisfy the relation $|\bm p_{\bar \pi}| = |\bm p_p| \cos \chi_1-|\bm p_{\bar p'}| \cos \epsilon_1$. 

Finally, the azimuthal, or Toller angle, $\omega$, is related to \mbox{$s = (p_{\overline{\gamma}}-p_p)^2$} via
\begin{align} \nonumber
 s &=  m_p^2 - \frac{1}{2t_{\gamma\eta}}(t_{\gamma\eta}-m_\eta^2)(t_{\gamma\eta}+m_p^2-s_{\pi p}) \\ \nonumber
&\quad  + \frac{1}{2t_{\gamma\eta}}\lambda^{1/2}(t_{\gamma\eta},0,m_\eta^2) \lambda^{1/2}(t_{\gamma\eta},m_p^2,s_{\pi p}) \\
&\quad \times \left( \sin\chi_1 \sin\theta_1 \cos\omega + \cos\chi_1 \cos\theta_1 \right) \ .
\label{eq:toller}
\end{align}

The $V_2$ rest frame corresponds to a boost of the two-proton system along $+{\bm {z}}$, with the $\pi^0$ crossed to the final state. The four-momenta are denoted $p'_p$, $p'_{\overline{p}'}$, $p'_{\overline{\gamma}}$, $p'_\eta$ and $p'_\pi$, and the energies $E'_p$, $E'_{\overline{p}'}$, $E'_{\overline{\gamma}}$, $E'_\eta$ and $E'_\pi$, and are defined analogously to the $V_1$ frame. 
We only quote the expression of the cosine of the polar angle $\theta_2$, related to $s_{\pi p} = (p'_{\pi} - p'_{\bar p'})^2$ by 
\begin{align}
   \cos\theta_2 & = - \frac{2 t_{pp} (s_{\pi p}-m_p^2-m_\pi^2) + t_{pp}(t_{pp}+m_\pi^2-t_{\gamma \eta})}
{ \lambda^{1/2}(t_{pp}, m_p^2, m_p^2) \lambda^{1/2}(t_{\gamma\eta},m_\pi^2,t_{pp})} \ .
\end{align}

It is worth mentioning that for fixed transferred momenta $t_{\gamma \eta}$ and $t_{pp}$, $\cos\theta_1$ and $\cos\theta_2$ are linear polynomials in $s_{\eta\pi}$ and $s_{\pi p}$, respectively. The combination \mbox{$z_\omega = \cos\theta_1 \cos\theta_2 \sin\omega$} is linear in $s$, or equivalently in $\kappa^{-1}$. 

\section{Derivation of the couplings}\label{sec:couplings}
The vector-vector-pseudoscalar vertex in~\cref{eq:VVP} leads to the following decay widths
\bsub\begin{align}\label{eq:WidthV}
\Gamma(V\to \gamma P) & = \frac{|g_{\gamma V P}|^2}{m_0^2}  \frac{p^3}{12\pi} \ , \\ \label{eq:WidthP}
\Gamma(P\to \gamma V) & = \frac{|g_{\gamma V P}|^2}{m_0^2} \frac{p^3}{4\pi} ,
\end{align}\esub
with the breakup momentum \mbox{$p=(m_V^2-m_P^2)/2m_V$} or \mbox{$p=(m_P^2-m_V^2)/2m_P$} for the respective decays. Using experimental meson masses and widths from the PDG~\cite{ParticleDataGroup:2024cfk}, we extract the relevant couplings from the radiative decays $\rho\to\gamma\pi^0,\gamma\eta$; $\omega\to\gamma\pi^0,\gamma\eta$; and $\eta'\to\gamma\rho^0,\gamma\omega$. 
The widths only determine the couplings squared $|g_{\gamma VP}|^2$. According to low-energy effective field theory, all these couplings are real and positive~\cite{Escribano:2005qq}. We thus chose the positive value of the square root.
Setting the reference scale $m_0=1\gev$, the results are summarized in~\cref{tab:couplings-all}~\cite{Mathieu:2017jjs}.

Vertices involving two vectors and one pseudoscalar, namely $g_{\omega\rho\pi}$, $g_{\rho\rho\eta^{(\prime)}}$, and $g_{\omega\omega\eta^{(\prime)}}$, require additional considerations, as these decays are not directly measurable. The $g_{\omega\rho\pi}$ coupling is estimated from the decay \mbox{$\omega\to\pi^+\pi^-\pi^0$}, saturated by the two-step process, \mbox{$\omega\to\rho\pi\to3\pi$}, following Ref.~\cite{Gell-Mann:1962hpq}. Although the first decay is kinematically forbidden for on-shell intermediate states (\ie $m_\omega<m_\rho+m_\pi$), the broad $\rho$ width allows the low-energy tail to contribute significantly to the decay amplitude. 

The decay $\omega\to\rho\pi$ is described by \mbox{$\frac{g_{\omega\rho\pi}}{m_0} i \epsilon_{\alpha \beta \mu\nu} \varepsilon^{\alpha*}_{\rho} q^\beta_{\rho} q_{\omega}^\mu \varepsilon^{\nu}_{\omega}$} [see~\cref{eq:VVP}] and the subsequent decay \mbox{$\rho\to \pi\pi$}, by $g_{\rho}\varepsilon_\rho^\mu(q_1-q_2)_\mu$, with $q_\rho=q_1+q_2$. Combining both, with $q_\rho=q_0+q_+$ and $q_\omega=q_0+q_++q_-$, and using the antisymmetry of the Levi-Civita tensor, the amplitude for \mbox{$\omega\to\rho^+\pi^-\to\pi^+\pi^0\pi^-$} reads
\begin{align}\label{eq:omega3pi}
    \langle\pi^+\pi^0\pi^-|\omega\rangle&=\frac{2g_\rho g_{\omega\rho\pi}}{m_0}\frac{1}{D_\rho(s_{+0})}i\epsilon_{\alpha\beta\mu\nu}q_0^\alpha q_+^\beta q_-^\mu \varepsilon^\nu_\omega \ ,
\end{align}
where $s_{+0}=(q_++q_0)^2$ and $D_\rho(s)=m_\rho^2-s-im_\rho\Gamma_\rho(s)$, with
\begin{align}
    \Gamma_\rho(s)=\Gamma_\rho\frac{m_\rho}{\sqrt{s}}\left(\frac{s-4m_\pi^2}{m_\rho^2-4m_\pi^2}\right)^{3/2} \ .
\end{align}

The coupling $g_\rho$ is fixed from $\Gamma(\rho\to \pi\pi)$
\begin{align}
    \Gamma_\rho=\frac{1}{3}\frac{4g_\rho^2}{8\pi m_\rho^2}p_\rho^3 \ ,
\end{align}
with $p_\rho$ the breakup momentum. Using experimental values~\cite{ParticleDataGroup:2024cfk} gives $|g_\rho| = 5.82$.

Including all three charge combinations ($\rho^+ \pi^-$, $\rho^- \pi^+$ and $\rho^0 \pi^0$), the full $\omega \to 3\pi$ width is obtained by integrating over the Dalitz plot,
\begin{align} \label{eq:omega3pi-width}
   & \Gamma_{\omega\to 3\pi}=\left(\frac{g_{\omega\rho\pi}g_\rho}{m_0}\right)^2\frac{1}{(2\pi)^3}\frac{4}{32m_\omega^3}\int ds_{+0}ds_{-0}\\ \nonumber
    &\times \frac{\Phi(s_{+0},s_{-0},s_{+-})}{4}\frac{1}{3}\left|\frac{1 }{D(s_{+0})}+\frac{1 }{D(s_{-0})}+\frac{1 }{D(s_{+-})}\right|^2 \ ,
\end{align}
with the Kibble function $\Phi(s_{+0},s_{-0},s_{+-})=s_{+0}s_{+-}s_{-0}-m_\pi^2(m_\omega^2-m_\pi^2)^2$ and $s_{+-}+s_{+0}+s_{-0}=m_\omega^2+3m_\pi^2$. In the narrow-width limit of the $\rho$ and neglecting interference among charge channels, this expression reduces to~\cref{eq:WidthV}.

The integration limits follow from the Dalitz plot kinematics. For fixed $s_{+0}$,
\begin{align}
    s_{-0}^{\text{min/max}}=\frac{1}{2}(m_\omega^2+3m_\pi^2-s_{+0})\pm 2p_+(s_{+0})p_-(s_{+0}) \ ,
\end{align}
with 
\bsub\begin{align}
    p_+(s_{+0})& =\frac{\lambda^{1/2}(s_{+0},m_\pi^2,m_\pi^2)}{2\sqrt{s_{+0}}} \ , \\
    p_-(s_{+0})&=\frac{\lambda^{1/2}(s_{+0},m_\omega^2,m_\pi^2)}{2\sqrt{s_{+0}}} \ ,
\end{align}\esub
and $s_{+0}$ spanning $ (2m_\pi)^2<s_{+0}<(m_\omega - m_\pi)^2 $. 

Numerical evaluation with experimental masses and widths gives $|g_{\omega\rho\pi}|=14.27$. We note that these decays do not fix the overall sign of this and the photon couplings, but this ambiguity does not affect the observables considered in this work. In principle, the sign could be determined by matching with low-energy experimental data.

The couplings $g_{\rho\rho\eta^{(\prime)}}$ and $g_{\omega\omega\eta^{(\prime)}}$ are obtained from $g_{\omega\rho\pi}$ using isospin relations. 
In the $\textrm{SU}(3)_\textrm{F}$ basis, the couplings to the light-quark states ($\eta_q$ and $\omega_q$) and strange-quark states ($\eta_s$ and $\omega_s$) follow from symmetry arguments. We assume ideal $\omega-\phi$ mixing, \ie $\omega\equiv \omega_q$ and $\phi \equiv \omega_s$, for the vector states. The couplings to the physical pseudoscalar states $\eta$ and $\eta'$ follow from $\eta=\eta_q \cos\phi_\eta-\eta_s\sin\phi_\eta$ and $\eta'=\eta_q\sin\phi_\eta+\eta_s \cos\phi_\eta$, where $\phi_\eta=41.4^\circ$ characterizes the light-quark content of the $\eta$ meson~\cite{Mathieu:2010ss} (see~\cref{tab:couplings-all}).

For the nucleon vertex, $g_1$ and $g_2$ in~\cref{eq:VNN} are related to the $s$-channel nucleon helicity non-flip and flip couplings, respectively, in nucleon-nucleon elastic scattering. 
At high energies and forward angles, this process is well described by a Regge-pole model, allowing the phenomenological extraction of the Regge residues through fits to experimental data~\cite{Irving:1977ea}. Matching the high-energy limit of the nucleon-nucleon elastic scattering amplitudes to the corresponding Regge expressions gives
\bsub\begin{align}
    A^{++}_{++}&\simeq\frac{2g_1^2s}{m_V^2-t}=\beta_{++}\frac{s/s_0}{\alpha'(m_V^2-t)}\beta_{++} \ , \\
    A^{+-}_{+-}&\simeq\frac{2g_1^2s}{m_V^2-t}=\beta_{+-}\frac{s/s_0}{\alpha'(m_V^2-t)}\beta_{+-}  \ ,
\end{align}\esub
where $\{+,-\}$ denote the helicities of the nucleons at each vertex. Taking the characteristic energy scale $s_0=1/\alpha'$ yields\footnote{We also chose the positive sign for the nucleon couplings. In practice, only the overall sign of each diagram matters.}
\begin{align}
    g_1=\frac{\beta_{++}}{\sqrt{2}} \ , \qquad g_2=\frac{\beta_{+-}}{\sqrt{2}} \ .
\end{align}
Using the phenomenological Regge residues in Ref.~\cite{Irving:1977ea} gives the couplings listed in~\cref{tab:couplings-all}.

\section{Kinematic singularities}
\label{sec:kinematicsingularities}
In order to derive the kinematic singularity and its double-Regge limit, we only keep the relevant factors in \cref{eq:Kfactor-tchannel} and define the quantity
\begin{align} \label{eq:Tlam}
     T_{\lambda_1,\,\lambda_2} & = \lambda_1 \sum_{\xi=-1}^1(-\xi)e^{-i\xi\omega}d^1_{\lambda_1\xi}(\theta_1)\,d^1_{\xi\lambda_2}(\theta_2) \ ,
\end{align}
where $\lambda_1 = \lambda_\gamma = \pm1$ and $\lambda_2 = \lambda - \lambda'=-1,0,1$ are the net helicities at the top and bottom vertex, respectively.
At high energies, the amplitude must factorize and obey $T_{-1,\,\lambda_2}= T_{+1,\,\lambda_2}$ and $T_{1,\,-\lambda_2}= (-1)^{\lambda_2} T_{1,\,\lambda_2}$, when both exchanges have natural parity. We can thus isolate the kinematic singularities corresponding to the double exchange of natural-parity mesons through the following combinations:
\bsub\begin{align}
    \Delta_{1,0} & =\frac{1}{2}\left(T_{+1,0}+T_{-1,0}\right) \ ,
    \\
    \Delta_{1,1}& = \frac{1}{4}\left[ T_{+1,+1} + T_{-1,+1} - T_{+1,-1} - T_{-1,-1} \right]\ ,
\end{align}\esub
corresponding to nucleon helicity non-flip and flip transitions, respectively. The other combinations are subleading in the doulbe Regge limit.

For the double-vector exchange model, the insertion of \cref{eq:Tlam} leads to
\bsub\begin{align}
    \Delta_{1,0} & = 
    \frac{-i}{\sqrt{2}} \cos\theta_1 \sin\theta_2 \sin\omega\, ,
    \\
    \Delta_{1,1}& = 
    \frac{i}{2}
    \cos\theta_1 \cos\theta_2 \sin\omega\, .
\end{align}\esub
Finally, using the kinematic relations in \cref{sec:kinematics}, the double-Regge limits take the form
\bsub\begin{align}
    \Delta_{1, 0} & \to \frac{1}{\sqrt{2}}\frac{s_1}{2|\bm p_1||\bm p_2|} \frac{s_2}{2|\bm p_2'||\bm p_p'|}\sin\omega \, ,
    \\
    \Delta_{ 1, 1} & \to \frac{- i}{2}\frac{s_1}{2|\bm p_1||\bm p_2|} \frac{s_2}{2|\bm p_2'||\bm p_p'|}\sin\omega \, .
\end{align}\esub

\section{Forward-backward asymmetry with kinematical cuts}\label{sec:asym-cuts}
Here, we compute the forward-backward asymmetry~\cref{eq:asym} after imposing kinematical cuts in the calculation of the forward and backward differential cross sections, designed to reproduce the effect of the cuts used in experimental analyses to suppress contributions from baryon-resonance backgrounds. The kinematical cuts were chosen to match ongoing GlueX analysis of $\eta\pi$ production in the double-Regge region~\cite{RebecaPhD}. To isolate the forward region, a cut is applied to $t_{\gamma\eta}$, whereas in the backward region it is applied to $t_{\gamma\pi}$,
\bsub
\begin{align}  \nonumber
F^{\rm cut}(s_{\eta\pi}) &= \int \diff^2\Omega_{\text{GJ}} \diff t_{pp} \frac{\diff ^4\sigma}{\diff^2\Omega_{\text{GJ}} \diff t_{pp} \diff s_{\eta\pi}}\ \\
&\times \theta(t_{\gamma\eta} > -2 \gevsq) \ \Theta(t_{pp}, s_{\eta\pi}, s_{\pi p}, s_{\eta p}) \ ,\\
\nonumber
B^{\rm cut}(s_{\eta\pi}) &= \int \diff^2\Omega_{\text{GJ}} \diff t_{pp} \frac{\diff ^4\sigma}{\diff^2\Omega_{\text{GJ}} \diff t_{pp} \diff s_{\eta\pi}}\\
& \times \theta(t_{\gamma\pi} > -2 \gevsq) \ \Theta(t_{pp}, s_{\eta\pi}, s_{\pi p}, s_{\eta p}) \ ,
\end{align} \esub
with the additional selection of the region of integration defined by
\begin{align} \nonumber
    &\Theta(t_{pp}, s_{\eta\pi}, s_{\pi p}, s_{\eta p})  = 
    \theta(t_{pp} > -2 \gevsq) \ \theta(s_{\eta\pi } > 4 \gevsq) 
    \\
    &\qquad \times \theta(s_{\pi p} > 4 \gevsq)\  \theta(s_{\eta p} > 4.75 \gevsq) \ .
\end{align}

\begin{figure}[htb]
\begin{center}
\includegraphics[width=\linewidth]{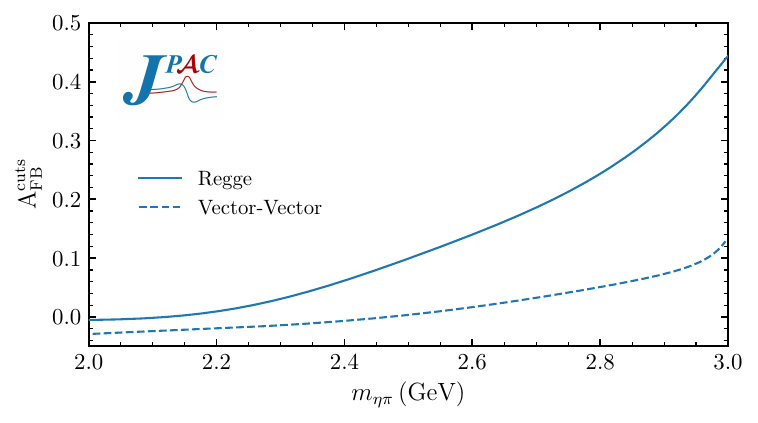}
\end{center}
\caption{Forward-backward asymmetry as a function of $\eta\pi$ invariant mass, including cuts consistent with the GlueX analysis strategy.}
\label{fig:asym-cuts}
\end{figure}

A comparison of the forward-backward asymmetry for $\eta\pi$ in~\cref{fig:asym-cuts} with that in~\cref{fig:asym} demonstrates that the inclusion of kinematical cuts has a significant impact on the asymmetry. In particular, the cuts applied here introduce a stronger dependence of the asymmetry as a function of $m_{\eta\pi}$.


\bibliographystyle{elsarticle-num} 

\bibliography{refs}
\end{document}